\documentclass[preprint2]{aastex6}  

\usepackage{natbib}
\usepackage{multirow}
\usepackage{color}
\usepackage{amsmath}
\usepackage{graphicx}
\usepackage{threeparttable}

\newcommand{\chandra}{\emph{Chandra}}
\newcommand{\rosat}{\emph{ROSAT}}

\newcommand{\jvla}{\emph{JVLA}}

\definecolor{green}{RGB}{102,153,102}
\definecolor{darkred}{RGB}{153,0,51}
\definecolor{blue}{RGB}{51,102,153}

\begin{document}


\title{Chandra and JVLA observations of HST Frontier Fields cluster MACS~J0717.5+3745}
\shorttitle{Chandra JVLA observations of MACS~J0717.5+3745}
\shortauthors{van Weeren et al.}

\author{R.~J.~van~Weeren\altaffilmark{1}$^\star$,  G.~A. Ogrean\altaffilmark{2}$^\dagger$, C.~Jones\altaffilmark{1}, W.~R.~Forman\altaffilmark{1},  F.~Andrade-Santos\altaffilmark{1}, Connor~J.~J.~Pearce\altaffilmark{3,1}, A.~Bonafede\altaffilmark{4}, M.~Br\"uggen\altaffilmark{4}, E.~Bulbul\altaffilmark{5}, T.~E.~Clarke\altaffilmark{6}, E.~Churazov\altaffilmark{7,8}, L.~David\altaffilmark{1}, W.~A.~Dawson\altaffilmark{9}, M.~Donahue\altaffilmark{10}, A.~Goulding\altaffilmark{11},  R.~P.~Kraft\altaffilmark{1}, B.~Mason\altaffilmark{12}, J.~Merten\altaffilmark{13}, T.~Mroczkowski\altaffilmark{14},  P.~E.~J.~Nulsen\altaffilmark{1,15}, P.~Rosati\altaffilmark{16}, E.~Roediger\altaffilmark{17}$^{\ddag}$,   S.~W.~Randall\altaffilmark{1}, J.~Sayers\altaffilmark{18},  K.~Umetsu\altaffilmark{19}, A.~Vikhlinin\altaffilmark{1}, A.~Zitrin\altaffilmark{18}$^\dagger$}

\email{rvanweeren@cfa.harvard.edu}

\altaffiltext{1}{Harvard-Smithsonian Center for Astrophysics, 60 Garden Street, Cambridge, MA 02138, USA}
\altaffiltext{2}{Department of Physics, Stanford University, 382 Via Pueblo Mall, Stanford, CA 94305-4060, USA}
\altaffiltext{3}{Department of Physics and Astronomy, University of Southampton, Southampton SO17 1BJ, UK}
\altaffiltext{4}{Hamburger Sternwarte, Universit\"at Hamburg, Gojenbergsweg 112, D-21029 Hamburg, Germany}
\altaffiltext{5}{Kavli Institute for Astrophysics and Space Research, Massachusetts Institute of Technology, \\77 Massachusetts Avenue, Cambridge, MA 02139}
\altaffiltext{6}{U.S. Naval Research Laboratory, 4555 Overlook Ave SW, Washington, D.C. 20375, USA}
\altaffiltext{7}{Max Planck Institute for Astrophysics, Karl-Schwarzschild-Str. 1, 85741, Garching, Germany}
\altaffiltext{8}{Space Research Institute, Profsoyuznaya 84/32, Moscow, 117997, Russia}
\altaffiltext{9}{Lawrence Livermore National Lab, 7000 East Avenue, Livermore, CA 94550, USA}
\altaffiltext{10}{Department of Physics and Astronomy, Michigan State University, East Lansing, MI 48824, USA}
\altaffiltext{11}{Department of Astrophysical Sciences, Princeton University, Princeton, NJ 08544, USA}
\altaffiltext{12}{National Radio Astronomy Observatory, 520 Edgemont Road, Charlottesville, VA 22903, USA}
\altaffiltext{13}{Department of Physics, University of Oxford, Keble Road, Oxford OX1 3RH, UK}
\altaffiltext{14}{ESO - European Organization for Astronomical Research in the Southern hemisphere, \\Karl-Schwarzschild-Str. 2, D-85748 Garching b. M\"unchen, Germany}
\altaffiltext{15}{ICRAR, University of Western Australia, 35 Stirling Hwy, Crawley WA 6009, Australia}
\altaffiltext{16}{Dipartimento di Fisica e Scienze della Terra, Universit\`a di Ferrara, Via Saragat 1, I-44122 Ferrara, Italy}
\altaffiltext{17}{E.A. Milne Centre for Astrophysics, Department of Physics \& Mathematics, \\University of Hull, Cottinton Road, Hull, HU6 7RX, UK}
\altaffiltext{18}{Cahill Center for Astronomy and Astrophysics, California Institute of Technology, MC 249-17, Pasadena, CA 91125, USA}
\altaffiltext{19}{Institute of Astronomy and Astrophysics, Academia Sinica, PO Box 23-141, Taipei 10617, Taiwan \\ \\ 
$\star$ Clay Fellow; $\dagger$ Hubble Fellow; $\ddag$ Visiting Scientist}



\begin{abstract}
To investigate the relationship between thermal and non-thermal components in merger galaxy clusters, we present deep \jvla\ and \chandra\ observations of the HST Frontier Fields cluster \object{MACS~J0717.5+3745}.   The \chandra\ image shows a complex merger event, with at least four components belonging to different merging subclusters. NW of the cluster, $\sim 0.7$~Mpc from the center, there is a  ram-pressure-stripped core that appears to have traversed the densest parts of the cluster after entering the ICM from the direction of a galaxy filament to the SE. We detect a density discontinuity NNE of this core which we speculate is associated with a cold front. Our radio images  reveal new details for the complex radio relic and radio halo in this cluster. In addition,  we discover several new filamentary radio sources with sizes of 100--300~kpc. A few of these seem to be connected to the main radio relic, while others are either embedded within the radio halo or projected onto it. A narrow-angled-tailed (NAT) radio galaxy,  a cluster member, is located at the center of the radio relic. The steep spectrum tails of this AGN leads into the large radio relic where the radio spectrum flattens again. This morphological connection between the NAT radio galaxy and relic provides evidence for re-acceleration (revival) of fossil electrons. The presence of hot $\gtrsim 20$~keV ICM gas detected by \chandra\   near the relic location provides additional support for this re-acceleration scenario.

\vspace{5mm}
\end{abstract}




\section{Introduction}

%
%
%

Merging galaxy clusters are excellent laboratories to investigate cluster formation and to explore how the particles that produce cluster-scale diffuse radio emission are accelerated. A textbook example of an extreme merging cluster is MACS~J0717.5+3745. MACS~J0717.5+3745 was discovered by \citet{2003MNRAS.339..913E} as part of the MAssive Cluster Survey \citep[MACS;][]{2001ApJ...553..668E} and is located at a redshift of $z=0.5458$. The cluster is very hot, with a global X-ray temperature of $11.6 \pm 0.5$~keV \citep{2007ApJ...661L..33E}. MACS~J0717.5+3745  is one of the most complex and dynamically disturbed clusters known. The cluster consists of at least four separate merging substructures. Regions of the intracluster medium (ICM) are heated to $\gtrsim 20$~keV \citep{2008ApJ...684..160M,2009ApJ...693L..56M,2012A&A...544A..71L,2016A&A...588A..99L}. A study of the Sunyaev-Zel'dovich (SZ) effect provided further evidence for the presence of shock-heated gas \citep{2012ApJ...761...47M}. \citet{2012ApJ...761...47M} and \citet{2013ApJ...778...52S} found evidence for a kinetic SZ signal for one of the subclusters, confirming the large velocity offset ($\approx 3000$~km~s$^{-1}$) found earlier from spectroscopy data \citep{2009ApJ...693L..56M}. A recent study reported the detection of a second kinetic SZ component belonging to another subcluster \citep{2016arXiv160607721A}. Connected to the cluster in the southeast is a $\sim4$~Mpc (projected length) galaxy and gas filament \citep{2004ApJ...609L..49E, 2012MNRAS.426.3369J}. 

Because of the large total mass  \citep[$M_{\rm{vir}} = \left(3.5 \pm 0.6\right) \times 10^{15} M_\odot$;][]{2014ApJ...795..163U}, complex mass distribution and relatively shallow mass profile \citep{2009ApJ...707L.102Z}, the cluster provides a large area of sky
with high lensing magnification, and is thus selected as part of the
Cluster Lensing And Supernova survey with Hubble \citep[CLASH,][]{2012ApJS..199...25P,2013ApJ...777...43M} and the HST Frontier Fields program\footnote{http://www.stsci.edu/hst/campaigns/frontier-fields/} \citep{2014AAS...22325401L,2016arXiv160506567L} to find high-z lensed objects.

\citet{2009ApJ...693L..56M} reported decrements in the ICM temperature near two of the subclusters of MACS~J0717.5+3745, which they interpret as remnants of cool cores. For one of these subclusters, \citet{2009ApJ...693L..56M} measured a temperature 5.7~keV temperature, suggesting this component is still at the early stage of merging. \citet{2015MNRAS.451.3920D} found that one of the dark matter components (the one furthest to the NW) has a significant offset from  the closest X-ray peak. Significant offsets between the lensing and X-ray peaks are expected in the case of a high-speed collision in the plane of the sky.

Previous radio studies of the cluster have focused on diffuse radio emission that is present in the cluster \citep{2009A&A...505..991V, 2009A&A...503..707B, 2013A&A...557A.117P}. The cluster hosts a giant radio halo extending over an {area of about 1.6~Mpc}. Polarized emission from the radio halo was detected by \citet{2009A&A...503..707B}. The radio luminosity (1.4~GHz radio power) is the largest known for any cluster, in agreement with the cluster's large mass and high global temperature \citep[e.g.,][]{2013ApJ...777..141C}. 

The cluster also hosts a  large 0.7--0.8~Mpc radio relic. It has been suggested that the radio relic in the cluster traces a large-scale shock wave which originated from the ongoing merger events \citep{2009A&A...505..991V}, or alternatively, from an accretion shock related to the large-scale filament at the southeast \citep{2009A&A...503..707B}. 

 In the standard scenario \citep{1998A&A...332..395E} for radio relics, particles are accelerated at the shock via the Diffusive Shock Acceleration (DSA) mechanism in a first order Fermi process \cite[e.g.][]{ 1983RPPh...46..973D}. A problem with this interpretation is that shock Mach numbers in clusters are typically low ($\mathcal{M} \lesssim 3$), in which case DSA is thought to be inefficient. 
For that reason several alternative models have been proposed including shock re-acceleration  \citep[e.g.,][]{2005ApJ...627..733M,2008A&A...486..347G,2011ApJ...734...18K,2012ApJ...756...97K, 2013MNRAS.435.1061P,a3411} and  turbulent re-acceleration \citep{2015ApJ...815..116F}. Recent work from particle in cell (PIC)  simulations  indicates that cluster shocks can inject electrons from the thermal pool \citep{2014ApJ...794..153G,2014ApJ...797...47G}, and that these electrons gain energy via the shock drift acceleration (SDA) mechanism.

In \citet{2016ApJ...817...98V}, we presented Karl G. Jansky Very Large Array (\jvla) and \chandra\ observations of lensed radio and X-ray sources located behind MACS~J0717.5+3745. In this work, we present the results of the \chandra\ and \jvla\ observations of the cluster itself. A \chandra\ analysis of the large-scale filament to the southeast is described in a separate letter \citep{gogreanfilament}.
The data reduction and observations are described in Section~\ref{sec:obs}. The radio and X-ray images, and the spectral index and ICM temperature maps are presented Sections~\ref{sec:radioresults} and~\ref{sec:xrayresults}. This is followed by a discussion and conclusions in Sections~\ref{sec:discussion} and~\ref{sec:conclusions}. In this paper we adopt a $\Lambda$CDM cosmology with $H_{\rm 0} = 70$~km~s$^{-1}$~Mpc$^{-1}$, $\Omega_{\rm m} = 0.3$, and $\Omega_{\Lambda} = 0.7$. With the adopted cosmology, 1\arcsec~corresponds to a physical scale of 6.387~kpc at $z=0.5458$. All our images are in the J2000 coordinate system.

\section{Observations \& data reduction}
\label{sec:obs}

\subsection{JVLA observations}
\jvla\ observations of MACS~J07175+3745 were obtained in the L-, S-, and C-bands, covering the frequency range from 1 to 6.5~GHz. An overview of the frequency bands and observations is given in Table~\ref{tab:jvlaobs}.  The total recorded bandwidth was 1~GHz for the L-band, and 2~GHz for the S- and C-bands. For the primary calibrators we used 3C138 and 3C147. J0713+4349 was included as a secondary calibrator. All four polarization products were recorded. 

The data were reduced with {\tt CASA} \citep{2007ASPC..376..127M} version 4.5 and data from the different observing runs were all processed in the same way. The data reduction procedure is described in more detail in \citet{2016ApJ...817...98V}. To summarize, the data were calibrated for the antenna position offsets, elevation dependent gains, global delay, cross-hand delay, bandpass, polarization leakage and angles, and temporal gain variations using the primary and secondary calibrator sources. RFI was identified and flagged with the {\tt AOFlagger} \citep{2010MNRAS.405..155O}. The calibration solutions from the primary and secondary calibrator sources were applied to the target field and several rounds of self-calibration were carried out to refine the gain solutions.

After the individual datasets were calibrated, the observations from the different configurations (for the same frequency band) were combined and imaged together. One extra round of self-calibration was  carried out, using the combined images, to align the datasets from the different configurations.

Deep continuum images were produced with {\tt WSClean} \citep{2014MNRAS.444..606O} in the three different frequency bands. We employed the wide-band clean and multiscale algorithms. Clean masks were employed at all stages and made with the {\tt PyBDSM} source detection package \citep{2015ascl.soft02007M}. The final images were corrected for the primary beam attenuation using the beam models provided by {\tt CASA}.

 Images were made with different weighting schemes to emphasize different aspects of the radio emission. An overview of the image properties is given in Table~\ref{tab:jvlaimages}. We also produced deeper images by stacking the L-, S-, and C-band images (equal weights) after convolving them to a common resolution\footnote{The large change in the primary beam size prevents a simple joint deconvolution and would have required the computationally expensive 
wide-band A-Projection algorithm \citep{2013ApJ...770...91B}.}.

\begin{table*}[]
\begin{center}
\caption{JVLA Observations}
\begin{tabular}{lllllll}
\hline
\hline
&    Observation date & Frequency coverage & Channel width & Integration time & On source time$^a$   &  LAS$^b$ \\
&                                      & [GHz]                                  & [MHz]                              & [s]                            &      [hr] &  [\arcsec] \\
\hline
L-band A-array & 28 Mar, 2013  & 1--2     & 1&1&$5.0$&36\\
L-band B-array &  25 Nov, 2013 & 1--2     & 1&3&$5.0$&120\\
L-band C-array & 11 Nov, 2014   & 1--2     & 1&5&$3.25$&970\\
L-band D-array &  9 Aug, 2014 & 1--2     & 1&5&$2.25$&970\\
S-band A-array & 22 Feb, 2013  & 2--4     & 2&1&$5.0$&18\\
S-band B-array & Nov 5, 2013    & 2--4     & 2&3&$5.0$&58\\ 
S-band C-array & 20 Oct, 2014 & 2--4     & 2&5&$3.25$&490\\
S-band D-array &  3 Aug, 2014& 2--4     & 2&5&$3.25$&490\\ 
C-band B-array & Sep 30, 2013 & 4.5--6.5&2&3&$5.0$&29\\
C-band C-array & Oct 13, 2014 & 4.5--6.5&2&5&$3.25$&240\\
C-band D-array & Aug 2, 2014 & 4.5--6.5&2&5&$3.25$&240\\
%
\hline
\hline
\end{tabular}
\label{tab:jvlaobs}
\end{center}
$^{a}$ Quarter hour rounding\\
$^{b}$ Largest angular scale that can be recovered by these observations\\
\end{table*}

\begin{table}[th!]
\begin{center}
\setlength{\tabcolsep}{2pt}
\caption{JVLA image properties}
\begin{tabular}{llllll}
\hline
\hline
& frequency & resolution & weighting$^{a}$ & uv-taper & r.m.s. noise \\
& [GHz] & [\arcsec] & &[\arcsec] & [$\mu$Jy]\\
\hline
& 1--2 &$1.3\arcsec\times1.1\arcsec$ & Briggs  & -- & 5.1 \\
& 1--2 &$2.6\arcsec\times2.4\arcsec$ &Briggs   &2 &4.9\\
& 1--2 &$5.0\arcsec\times4.9\arcsec$ &uniform&5 &7.9\\
& 1--2 &$10.1\arcsec\times9.9\arcsec$ &uniform&10 & 15\\
& 2--4 &$0.85\arcsec\times0.59\arcsec$ & Briggs  & -- & 1.9\\
& 2--4 &$2.3\arcsec\times2.2\arcsec$ &Briggs   &2 & 2.0 \\
& 2--4 &$5.0\arcsec\times4.9$ &uniform&5 & 6.9\\
& 2--4 &$10.1\arcsec\times9.9\arcsec$ &uniform&10 & 6.2\\
& 4.5--6.5 &$1.7\arcsec\times1.2\arcsec$ & Briggs & -- & 2.2\\
& 4.5--6.5 &$3.0\arcsec\times2.5\arcsec$ &Briggs   &2 & 2.0\\
& 4.5--6.5 &$5.0\arcsec\times4.9\arcsec$ &uniform&5 & 2.4\\
& 4.5--6.5 &$10.1\arcsec\times9.9\arcsec$ &uniform&10 &3.9\\
\hline
\hline
\end{tabular}
\label{tab:jvlaimages}
\end{center}
$^{a}$ For all images made with \citet{briggs_phd} weighting we used {\tt robust=0}. \\
\end{table}

\subsubsection{Spectral index maps}
\label{sec:makespix}
For making radio spectral index maps, we imaged the data with {\tt CASA}, employing w-projection \citep{2005ASPC..347...86C,2008ISTSP...2..647C} and multiscale clean  \citep{2008ISTSP...2..793C} with {\tt nterms=3} \citep[][]{2011A&A...532A..71R}. 
Three separate continuum images (corresponding to the L-, S-, and C-bands), were created at reference frequencies of 1.5, 3.0, and 5.5~GHz, respectively.  Inner uv-range cuts were employed based on minimum uv-distance provided by the C-band data. In addition, we used uniform weighting to correct for differences in the uv-plane sampling. Different uv-tapers were used to produce images at resolutions of 1.5\arcsec, 2.5\arcsec, 5\arcsec~and 10\arcsec. The remaining minor differences in the beam sizes (after using the uv-tapers) were taken out by convolving the images to the same resolution. The images were  corrected for the primary beam attenuation. 

We  created the spectral index maps by fitting a first order polynomial through the three flux measurements at 1.5, 3.0 and 5.5~GHz in $\log{(S)}-\log{(\nu)}$ space. The spectral index thus represents the average spectral index  in the 1.0--6.5~GHz band, neglecting any spectral curvature. Pixels with values below $2.5\sigma_{\rm{rms}}$  in any of the three maps were blanked.

\subsection{Chandra Observations}
\label{sec:chandraobs}

\begin{table*}[h!]
\begin{center}
\caption{Summary of the  \emph{Chandra} observations.}
\begin{tabular}{lccccc}
\hline
\hline
ObsID & Instrument & Mode & Start date & Exposure time (ks) & Filtered exposure time (ks) \\
\hline
 1655$^a$ & ACIS-I &    FAINT & 2001-01-29 & 19.9 & 15.8  \\
 4200 & ACIS-I & VFAINT & 2004-01-10 & 59.0 & 52.6  \\
16235 & ACIS-I &    FAINT & 2013-12-16 & 70.2 & 63.4  \\
16305 & ACIS-I & VFAINT & 2013-12-11 & 94.3 & 82.6  \\
\hline
\hline
\end{tabular}
\label{tab:chandraobs}
\end{center}
$^a$ ObsID~1655 was excluded from the analysis, see Section~\ref{sec:chandraobs}
\end{table*}

MACS~J0717.5+3745 was observed with \chandra\ for a total of 243~ks between 2001 and 2013. A summary of the observations is presented in Table~\ref{tab:chandraobs}. The datasets were reduced with CIAO v4.7 and CALDB v4.6.5, following the same methodology that was described by \citet{2015ApJ...812..153O}. ObsID~1655 was contaminated by flares even after the standard cleaning was applied. Given that the exposure time of ObsID~1655 is $<10\%$ of the total exposure time, we decided to exclude this ObsID from the analysis.

The instrumental background was modeled using stowed background event files appropriate for the dates of the observations (period B event files for ObsID 4200, and period F event files for ObsIDs 16235 and 16305). The stowed background spectra and images were normalized to have the same $10-12$~keV count rate as the corresponding ObsID. 

Point sources were detected with the CIAO script {\tt wavdetect} using wavelet scales of 1, 2, 4, 8, 16, and 32~pixels and ellipses with radii $5\sigma$ around the centers of the detected sources. These point sources were excluded from the analysis.

\subsection{Chandra Background Modeling}

Background spectra were extracted from a region outside $2.5$~Mpc from the cluster center. The instrumental background was subtracted from them, and the sky background was modeled with the sum of an unaborbed thermal component \citep[APEC;][]{Smith2001} describing emission from the Local Hot Bubble (LHB), an absorbed thermal component describing emission from the Galactic Halo (GH), and an absorbed power-law component describing emission from unresolved point sources. We used photoelectric absorption cross-sections from \citet{Verner1996}, and the elemental abundances from \citet{Feldman1992}. The hydrogen column density in the direction of MACS~J0717.5+3745 was fixed to $8.4\times 10^{20}$~cm$^{-2}$, which is the sum of the weighted average atomic hydrogen column density from the Leiden-Argentine-Bonn \citep[LAB;][]{Kalberla2005} Survey and the molecular hydrogen column density determined by \citet{Willingale2013} from \emph{Swift} data. The temperatures and the normalizations of the thermal components were free in the fit, but linked between different datasets. The temperature and normalization of the LHB component are difficult to constrain from the \chandra\ data (its temperature is $\sim 0.1$~keV), so we determined them from a \rosat\ spectrum extracted from an annulus with radii $0.15$ and $1$ degrees around the cluster center (which is beyond $R_{\rm 200}$). The index of the power-law component was fixed to 1.41 \citep{DelucaMolendi2004}. The normalizations of the power-law components of the \chandra\ spectra were free in the fit, but the power-law normalizations of ObsIDs 16235 and 16305 were linked since the observations were taken close in time. The power-law normalization of the \rosat\ spectrum was fixed to $8.85\times 10^{-7}$ photons~keV$^{-1}$~cm$^{-2}$~s$^{-1}$~arcmin$^{-2}$ \citep{Moretti2003}. The instrumental background-subtracted spectra were modeled with {\tt xspec}\footnote{AtomDB version is 2.0.2} v12.8.2 \citep{Arnaud1996}. The \chandra\ spectra were binned to a minimum of 1 count/bin, and modeled using the extended C-statistic \citep{Cash79,1979ApJ...230..274W}. The spectra were fitted in the energy band $0.5-7$~keV.\footnote{The same energy band was used for all the spectral fits presented in this paper.} The best-fitting sky background parameters are summarized in Table~\ref{tab:skybkg}. Throughout the paper, the uncertainties for the X-ray derived quantities are quoted at the $1\sigma$ level, unless explicitly stated.

\begin{table*}
\setlength{\tabcolsep}{3pt}
\begin{center}
\caption{Best-fitting X-ray sky background parameters.}
\begin{tabular}{lclcc}
\hline
\hline
Model component & Parameter & \multicolumn{2}{c}{Chandra} & ROSAT \\
\hline
\multirow{2}{*}{{LHB}} & {$kT$ (keV)} & \multicolumn{2}{c}{{$0.135$ (fixed)}} &{$0.135_{-0.08}^{+0.07}$} \\
     & {norm (cm$^{-5}$~arcmin$^{-2}$)} & \multicolumn{2}{c}{{$7.21\times 10^{-7}$ (fixed)}} & {$7.21_{-0.18}^{+0.30}\times 10^{-7}$} \\
\hline
\multirow{2}{*}{{GH}} & {$kT$ (keV)} & \multicolumn{2}{c}{{$0.59_{-0.08}^{+0.09}$}} & {$0.64_{-0.28}^{+0.30}$} \\     
     & {norm (cm$^{-5}$~arcmin$^{-2}$)} & \multicolumn{2}{c}{{$2.79_{-0.44}^{+0.45}\times 10^{-7}$}} & {$3.79_{-0.85}^{+2.26}\times 10^{-7}$} \\ 
\hline
\multirow{4}{*}{{Power-law}} & {$\Gamma$} & \multicolumn{2}{c}{{$1.41$ (fixed)}} & {$1.41$ (fixed)} \\
     & \multirow{3}{*}{{norm at 1~keV (photons~keV$^{-1}$~cm$^{-2}$~s$^{-1}$~arcmin$^{-2}$)}} & {ObsID 4200} & {$7.00_{-0.56}^{+0.51} \times 10^{-7}$} & \multirow{3}{*}{{$8.85\times 10^{-7}$ (fixed)}}\\
     &   & {ObsIDs 16235} & \multirow{2}{*}{{$4.45^{+0.33}_{-0.39} \times 10^{-7}$}} &  \\
     &   & {ObsIDs 16305} &   &  \\
\hline\hline
\end{tabular}
\label{tab:skybkg}
\end{center}
\end{table*}

\section{Radio Results}
\label{sec:radioresults}

\subsection{Continuum images}
\begin{figure*}[h]
\begin{center}
\includegraphics[angle =180, trim =0cm 0cm 0cm 0cm,width=0.75\textwidth]{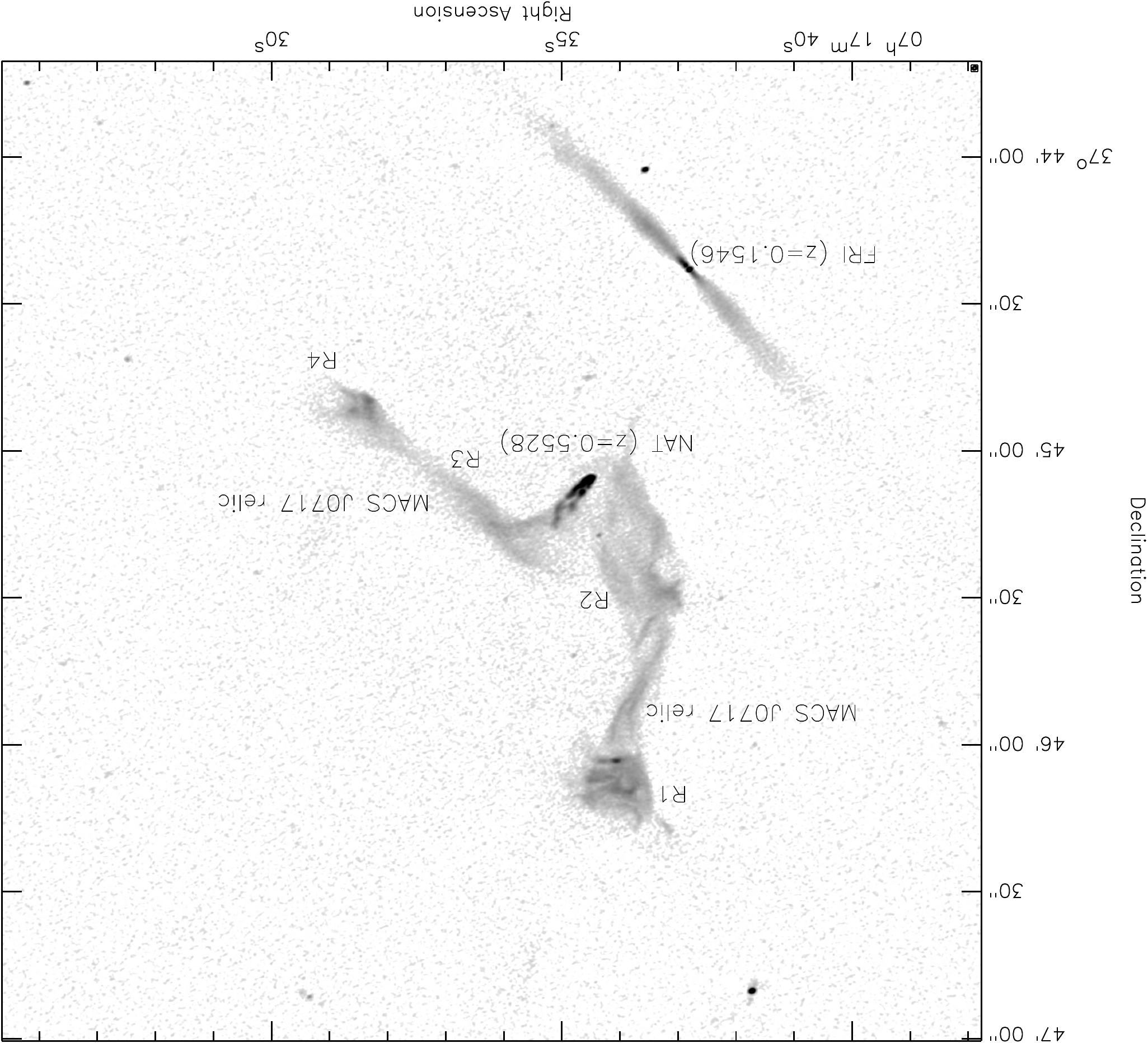}
\end{center}
\caption{S-band 2--4~GHz combined \jvla\ A-, B-, C-, and D-array image made with \citet{briggs_phd} weighting ({\tt robust=0}). The image has a resolution of $0.85\arcsec\times0.59\arcsec$  and a r.m.s. noise level of 1.9~$\mu$Jy~beam$^{-1}$. Components of the radio relic are labeled as in \citet{2009A&A...505..991V}.}
\label{fig:sband}
\end{figure*}

\begin{figure*}[h]
\centering
\includegraphics[angle =180, trim =0cm 0cm 0cm 0cm,width=1\columnwidth]{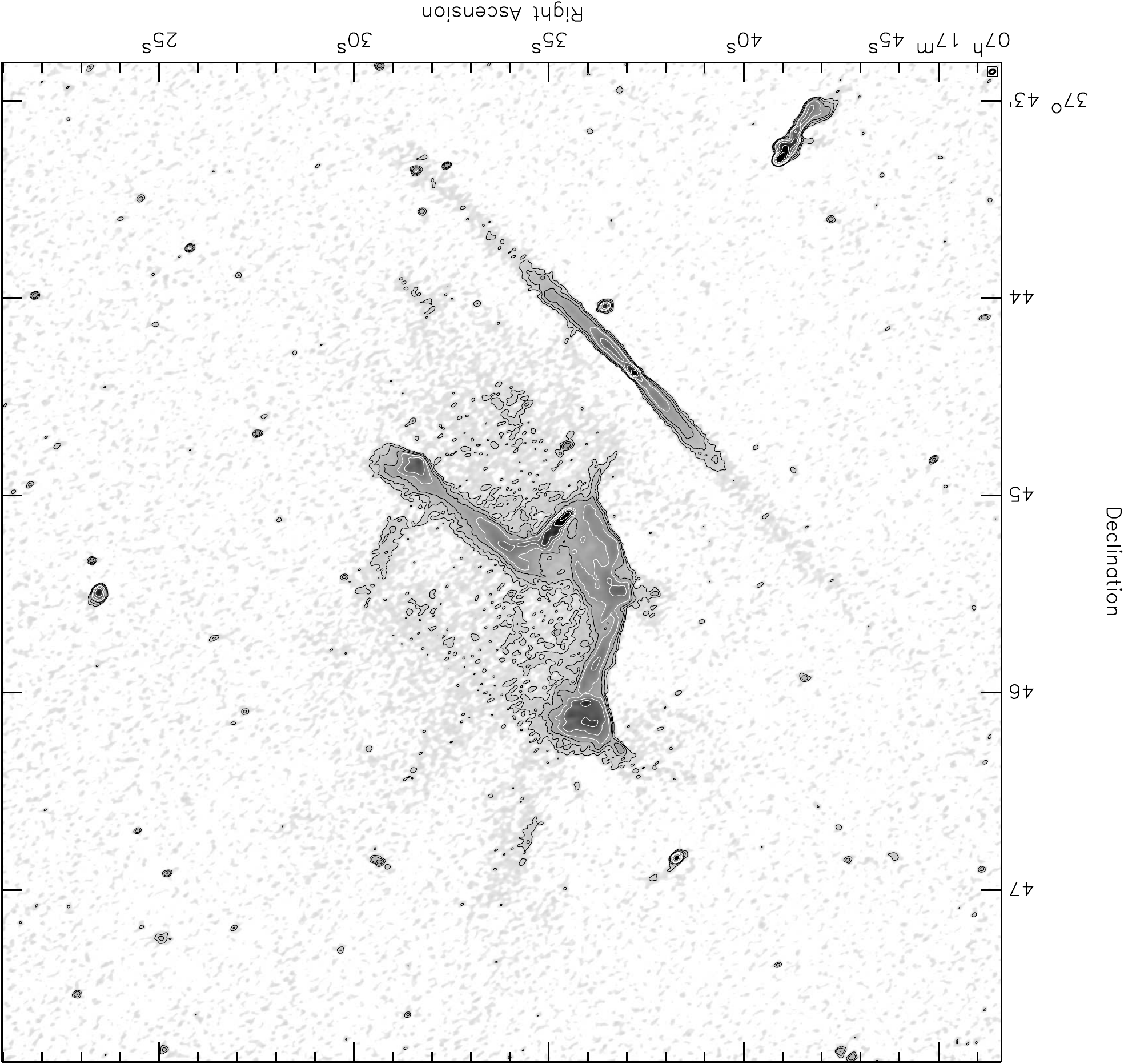}
\includegraphics[angle =180, trim =0cm 0cm 0cm 0cm,width=1\columnwidth]{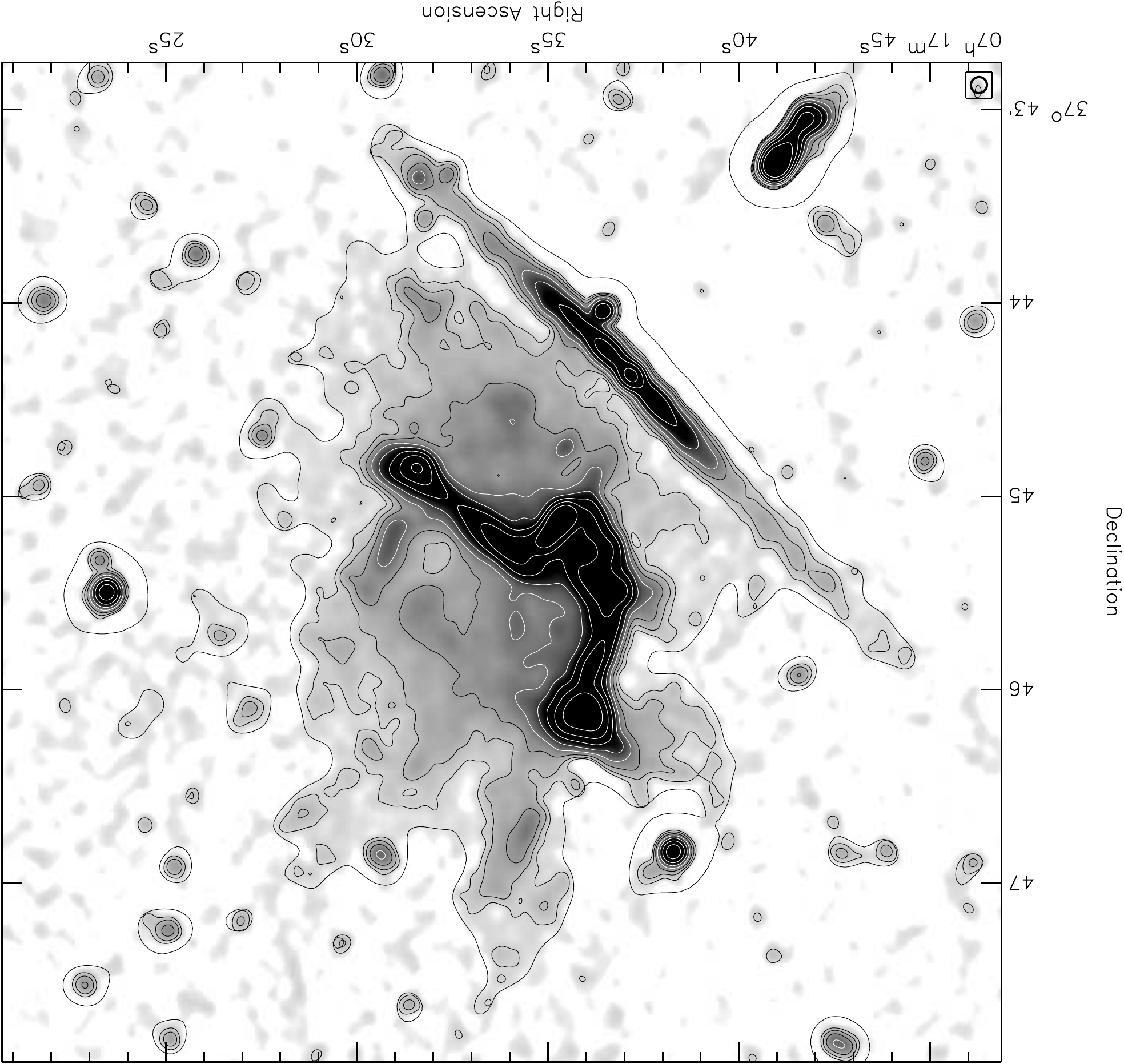}
\caption{Left: Deep wide-band combined  L-, S-, and C-band image with a resolution of 1.6\arcsec. Contour levels are drawn at $[1, 2, 4, \ldots] \times 4\sigma_{\rm{rms}}$. These individual  L-, S-, and C-band images were made with \citet{briggs_phd} weighting ({\tt robust=0}).
Right: Deep wide-band combined  L-, S-, and C-band image with a resolution of 5\arcsec. Contours are plotted in the same way as in the left panel, with the exception of the lowest contour level.  The lowest contour level comes from the 10\arcsec~resolution image and is drawn at $5\sigma_{\rm{rms}}$. The 5\arcsec~and 10\arcsec~resolution images were made with uniform weighting and tapered to these respective resolutions.}
\label{fig:deep16}
\end{figure*}

The high-resolution ($0.85\arcsec\times0.59\arcsec$) 2--4~GHz S-band image is shown in Figure~\ref{fig:sband}. Combined wide-band \mbox{L-,} S-, and C-band images at 1.6\arcsec~and 5\arcsec~resolution are shown in Figure~\ref{fig:deep16}. The most prominent source in the images is the large filamentary radio relic, with an embedded Narrow Angle Tail (NAT) galaxy \citep[$z=0.5528$,][]{2014ApJS..211...21E}  at the center of the main structure. The tails of the radio source are aligned with the radio relic.  Various components of the radio relic are labeled as in \citet{2009A&A...505..991V} on Figure~\ref{fig:sband}.

\begin{figure*}[h]
\begin{center}
\includegraphics[angle =0, trim =0cm 0cm 0cm 0cm,width=1.0\textwidth]{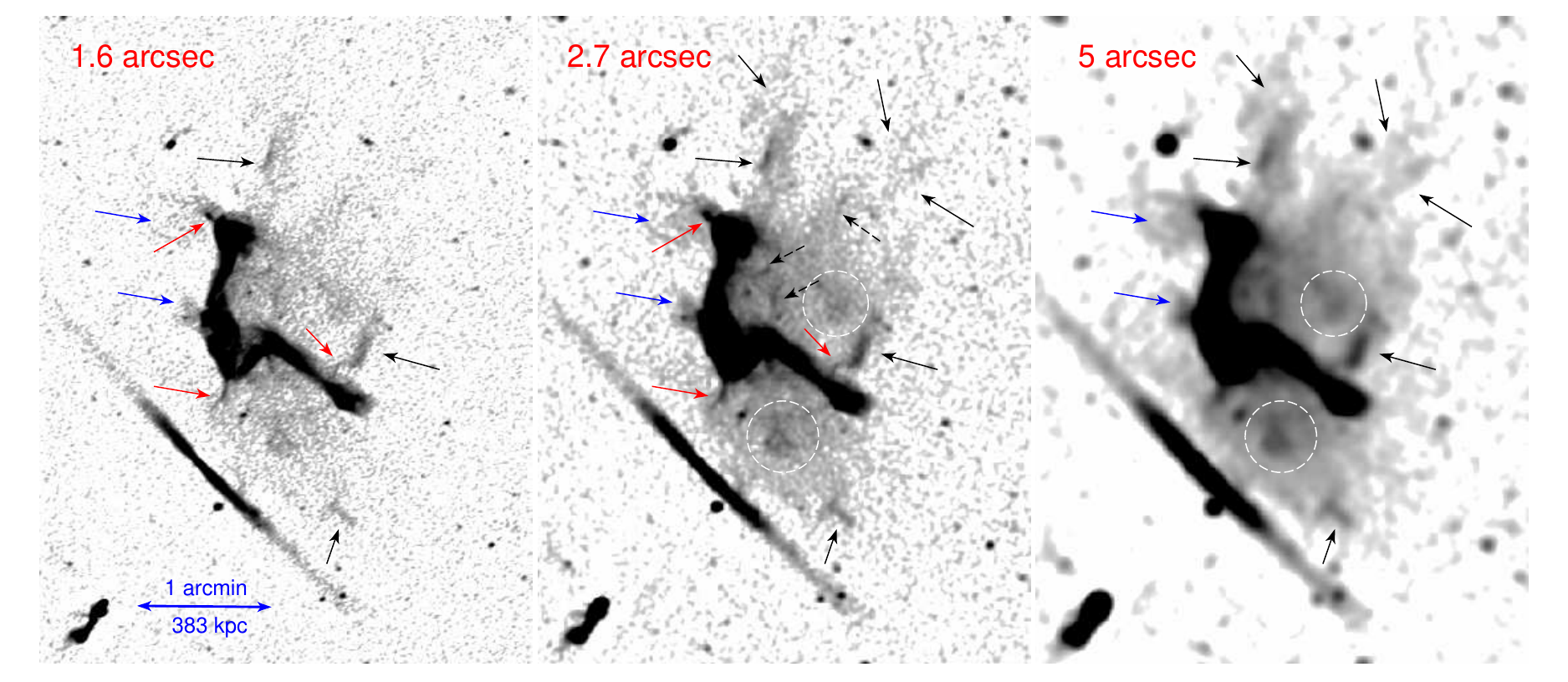}
\end{center}
\caption{Wide-band 1.0--6.5~GHz images of the cluster at resolutions of 1.6, 2.7 and 5\arcsec. The weighting schemes to make the 1.6 and  5\arcsec~resolution images is given in the caption of Figure~\ref{fig:deep16}. The 2.7\arcsec~resolution image was made with \citet{briggs_phd} weighting ({\tt robust=0}) and a uv-taper. These wide-band images reveal a significant amount of fine-scale structure in the extended radio emission. Narrow filaments extending from the relic are indicated with red arrows, diffuse components extending from the relic with blue arrows, and (small) filaments in the general region of the halo with black (dashed-line) arrows. Two enhancements in the radio halo emission are indicated with white dashed-line circles.}
\label{fig:3res}
\end{figure*}

A bright linearly shaped FRI radio source \citep{1974MNRAS.167P..31F} is located to the SE. This source is associated with an elliptical foreground galaxy (2MASX J07173724+3744224 ) located at z=0.1546 \citep{2009A&A...503..707B}. 
Another tailed radio source at the far SE is located at $z=0.5399$ \citep{2014ApJS..211...21E}, see Figure~\ref{fig:deep16}. This radio galaxy is probably falling into the cluster along the large-scale galaxy filament to the southeast \citep[e.g.,][]{2004ApJ...609L..49E}, given that the tails align with the galaxy filament and point to the southeast. The combined L-, S-, and C-band \jvla\ images at resolutions of 1.6\arcsec, 2.7\arcsec, and 5\arcsec -- to highlight details around the radio halo area -- are shown in Figure~\ref{fig:3res}.

For the relic, we measure a largest linear size (LLS) of $\approx800$~kpc, similar to previous studies. Our new images are significantly deeper and have better resolution than previous studies of this source. They reveal many new details in the relic and show that the relic has a significant amount of filamentary structure on scales down to $\sim30$~kpc. Small scale filamentary structures have also been seen for other relics, such as the Toothbrush cluster \citep{2012A&A...546A.124V,2016ApJ...818..204V}, A3667 \citep{1997MNRAS.290..577R}, A3376 \citep{2006Sci...314..791B} and in particular for Abell~2256 \citep{2006AJ....131.2900C,2009A&A...508.1269V,2014ApJ...794...24O}.

We also note several narrow filaments of emission originating from the relic. These are marked with red arrows in Figure~\ref{fig:3res}. These filaments have lengths of 50--150~kpc and widths as small as 10~kpc. In addition, there are two larger regions of extended emission that are connected to the radio relic. These extended regions are marked with blue arrows.

The radio halo component extends to the south and north of  R3 and west of the R2 (see Figure~\ref{fig:sband} for the labeling). Our images also reveal a significant amount of structure around the radio halo, including several filamentary features. They are marked with black arrows in Figure~\ref{fig:3res}. The brightest of these is connected with the R3 component of the main radio relic and has a LLS of about 200~kpc. 

Another prominent $\sim 200$~kpc NS elongated radio filament is located at the northern outskirts of the cluster. This filament has  a well defined boundary on its eastern side, while it fades gradually towards the west. A fainter filament with a similar size and NS orientation is located NE of it. Evidence of two other filaments, with a NW orientation, are seen at the northern boundary of the radio halo. Another $\sim 100$~kpc long structure is located at the southern end of the radio halo. Three additional embedded filamentary structures in the radio halo are found west of R2. These are marked with dashed-line black arrows ( Figure~\ref{fig:3res}). We also find two enhancements in the halo emission which we marked with white dashed-line circles. 

\citet{2009A&A...503..707B} suggested the presence  of faint radio emission to the SE, along the large-scale galaxy filament (not to be confused with the smaller radio filaments in the cluster) in the 325~MHz image from the WENSS survey \citep{1997A&AS..124..259R}. We do not find  evidence for this in our  deeper observations. We speculate that the emission seen at 325~MHz could have been the result of blending of several compact sources due to the low-resolution of the 325~MHz image. We also note that the emission is not found in the low-frequency GMRT 610 and 235~MHz observations published by \citet{2013A&A...557A.117P}.

\subsubsection{Spectral index maps}
\label{sec:spix}

\begin{figure*}[h!]
\begin{center}
\includegraphics[angle =180, trim =0cm 0cm 0cm 0cm,width=1\columnwidth]{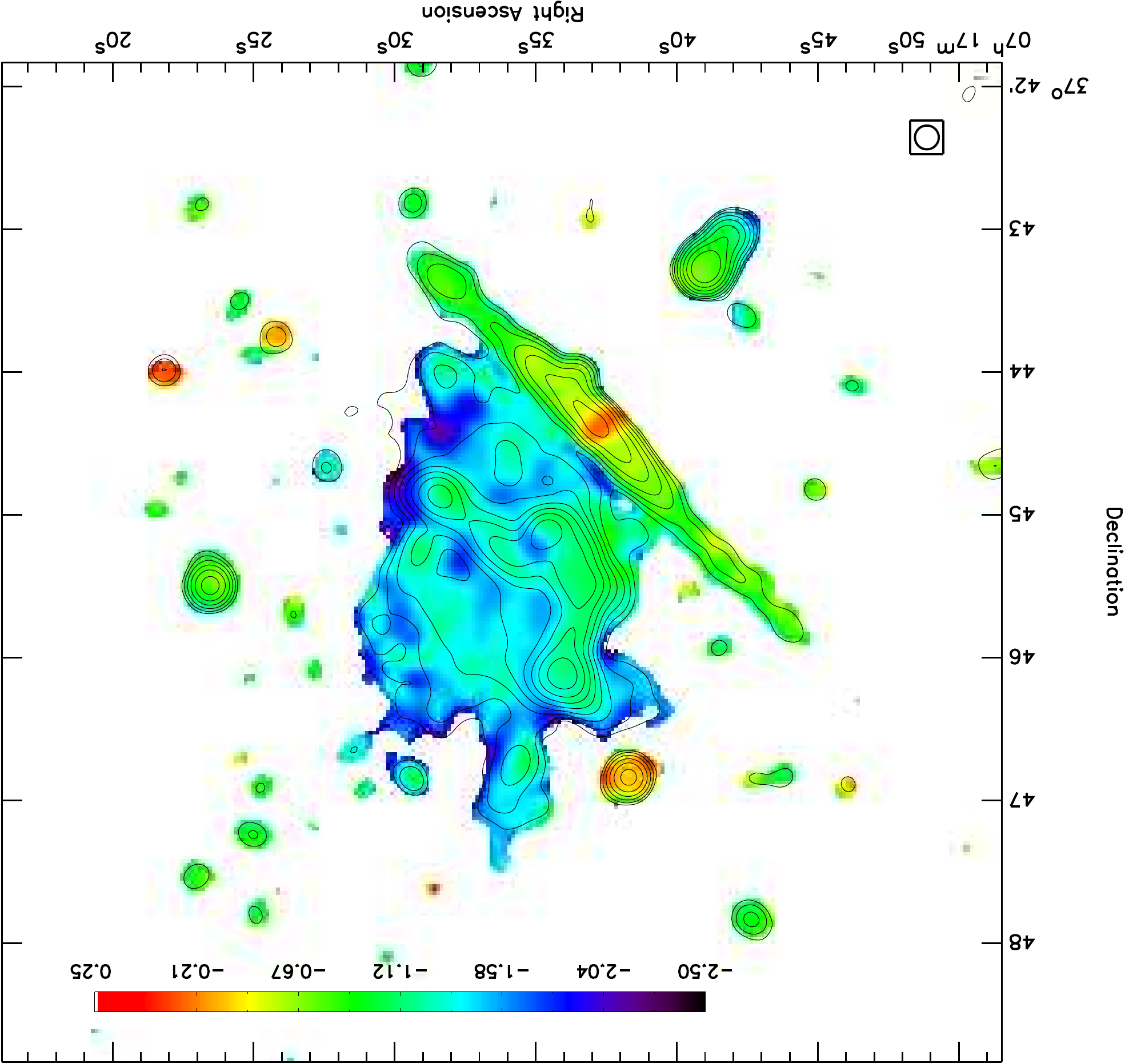}
\includegraphics[angle =180, trim =0cm 0cm 0cm 0cm,width=1\columnwidth]{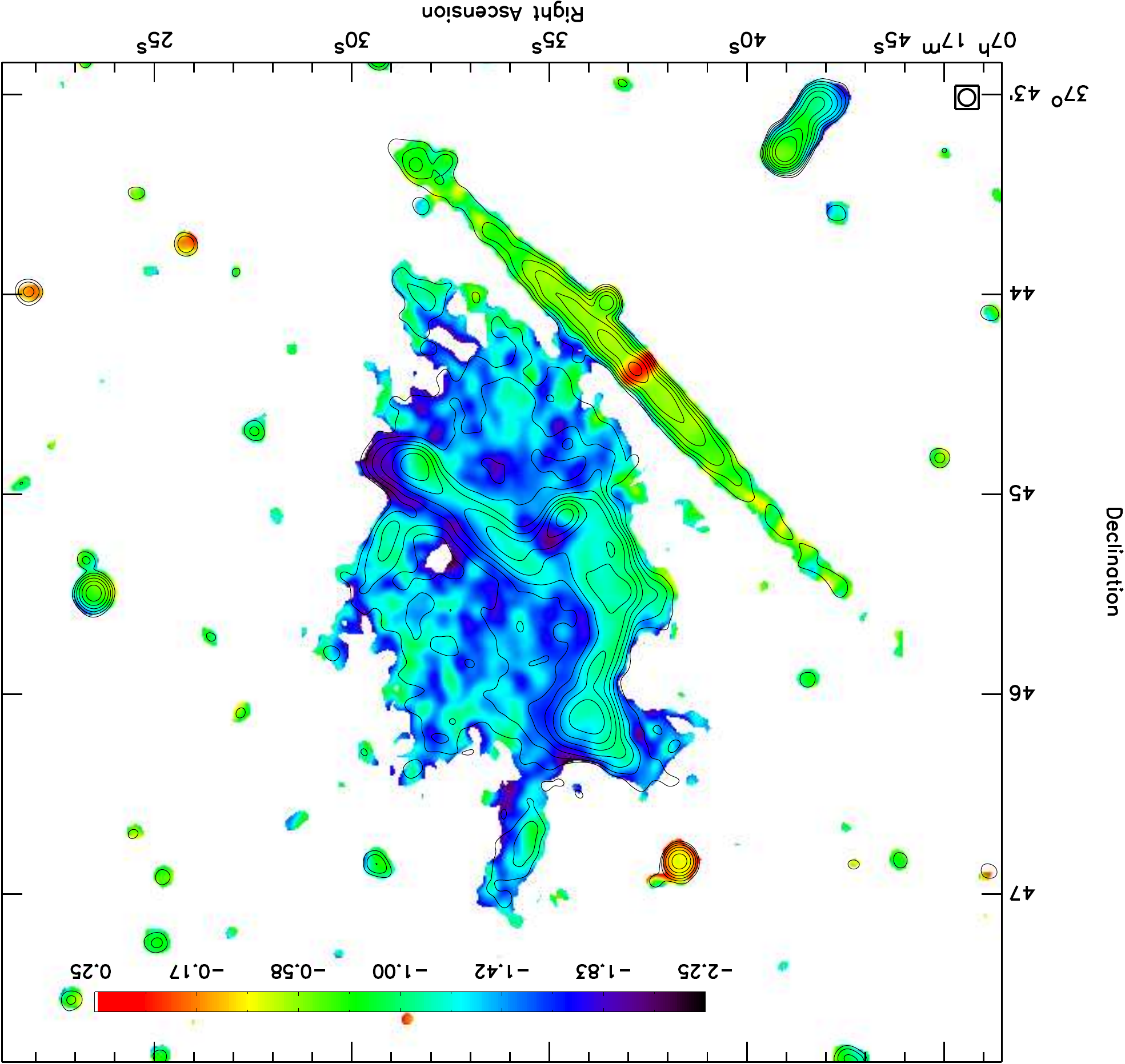}
\includegraphics[angle =180, trim =0cm 0cm 0cm 0cm,width=1\columnwidth]{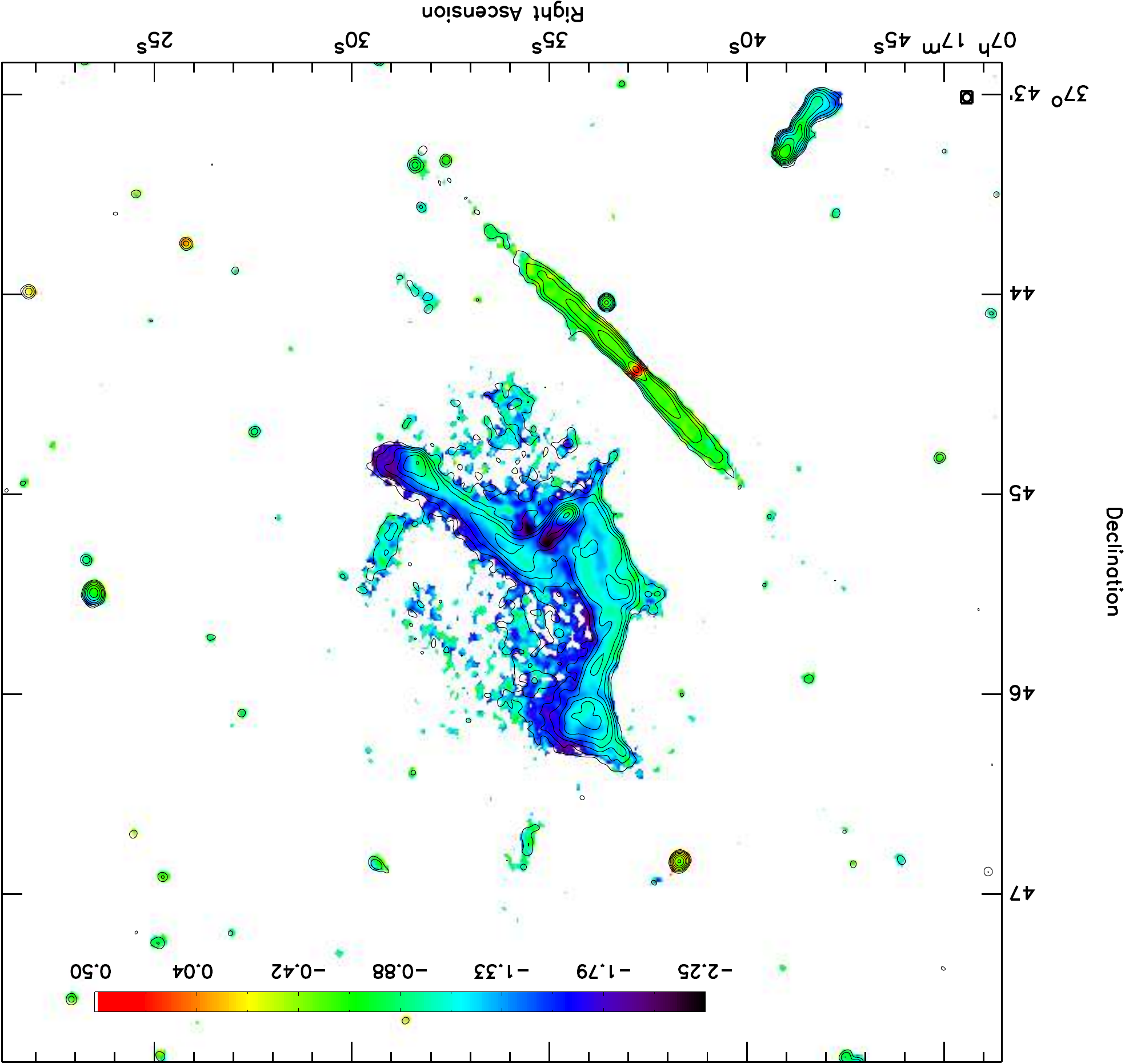}
\includegraphics[angle =180, trim =0cm 0cm 0cm 0cm,width=1\columnwidth]{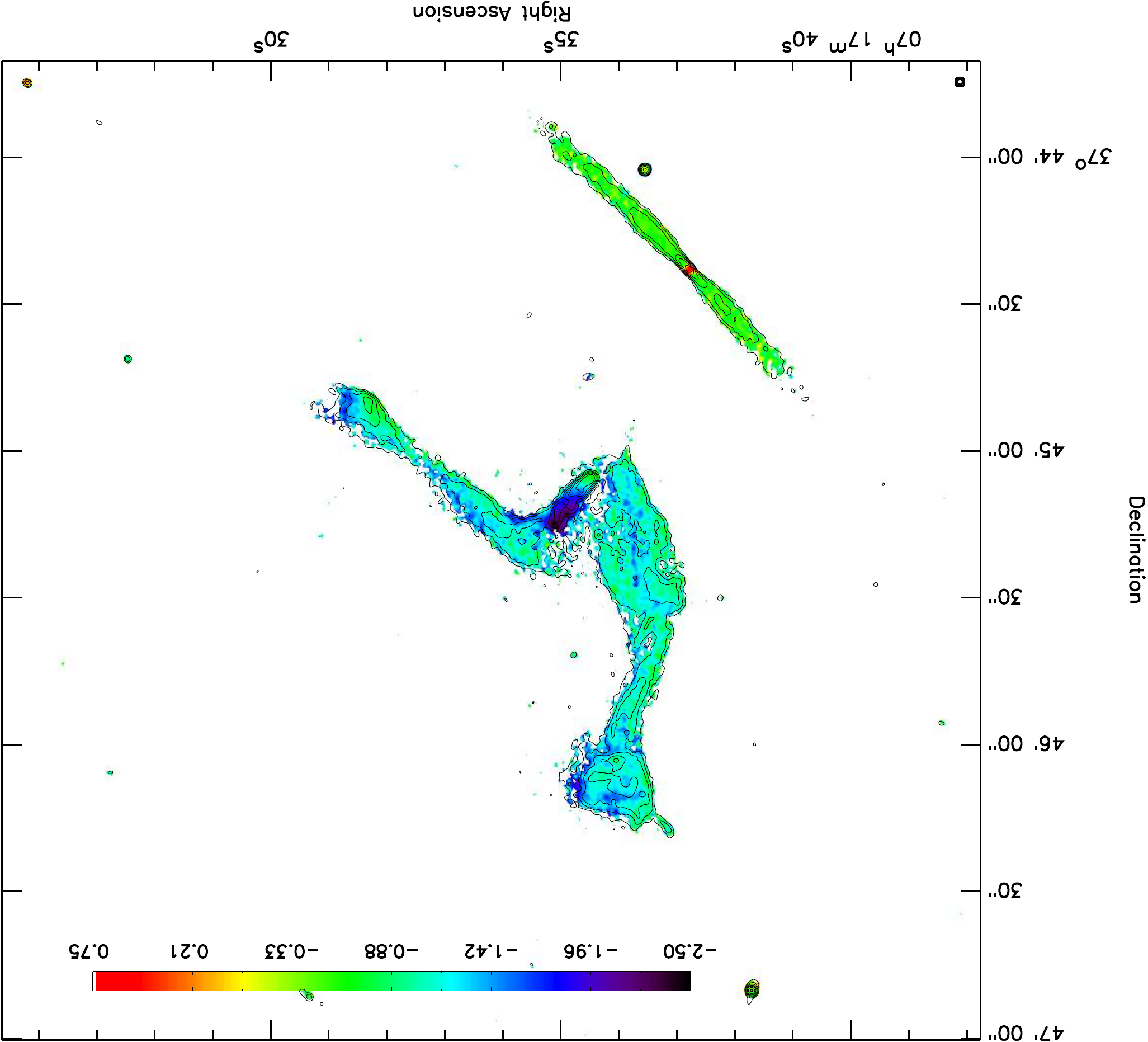}
\end{center}
\caption{Spectral index maps at 10\arcsec, 5\arcsec, 2.5\arcsec, and 1.2\arcsec~resolution (top left to bottom right). Black contours are drawn at levels of $[1,2,4,\ldots] \times 5\sigma_{\rm{rms}}$ and are from the S-band image. Pixels with values below $2.5\sigma_{\rm{rms}}$ in the individual maps were blanked. The corresponding spectral index uncertainty maps are displayed in Figure~\ref{fig:spixu}.}
\label{fig:spix}
\end{figure*}

Spectral index maps at resolutions of 10\arcsec, 5\arcsec, 2.5\arcsec, and 1.2\arcsec~are shown in Figure~\ref{fig:spix}. As explained in Section~\ref{sec:makespix}, these were made by fitting straight power-laws through flux measurements at 1.5, 3.0, and 5.5~GHz. The spectral index uncertainty maps are shown in Appendix~\ref{sec:spixerror}.

\begin{figure*}[h]
\begin{center}
\includegraphics[angle =0, trim =0cm 12cm 0cm 0cm,width=0.49\textwidth]{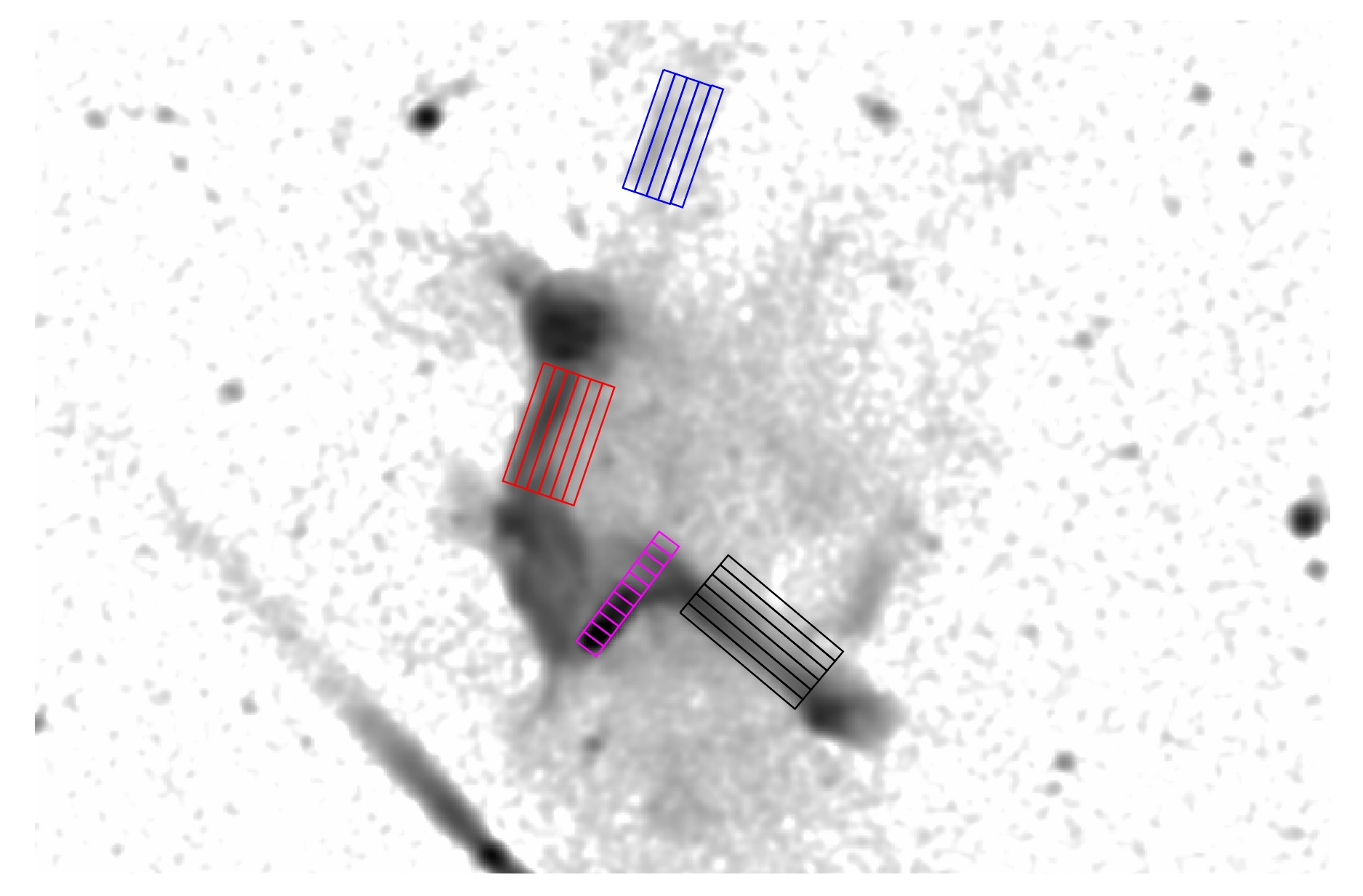}
\includegraphics[angle =180, trim =0cm 0cm 0cm 0cm,width=0.44\textwidth]{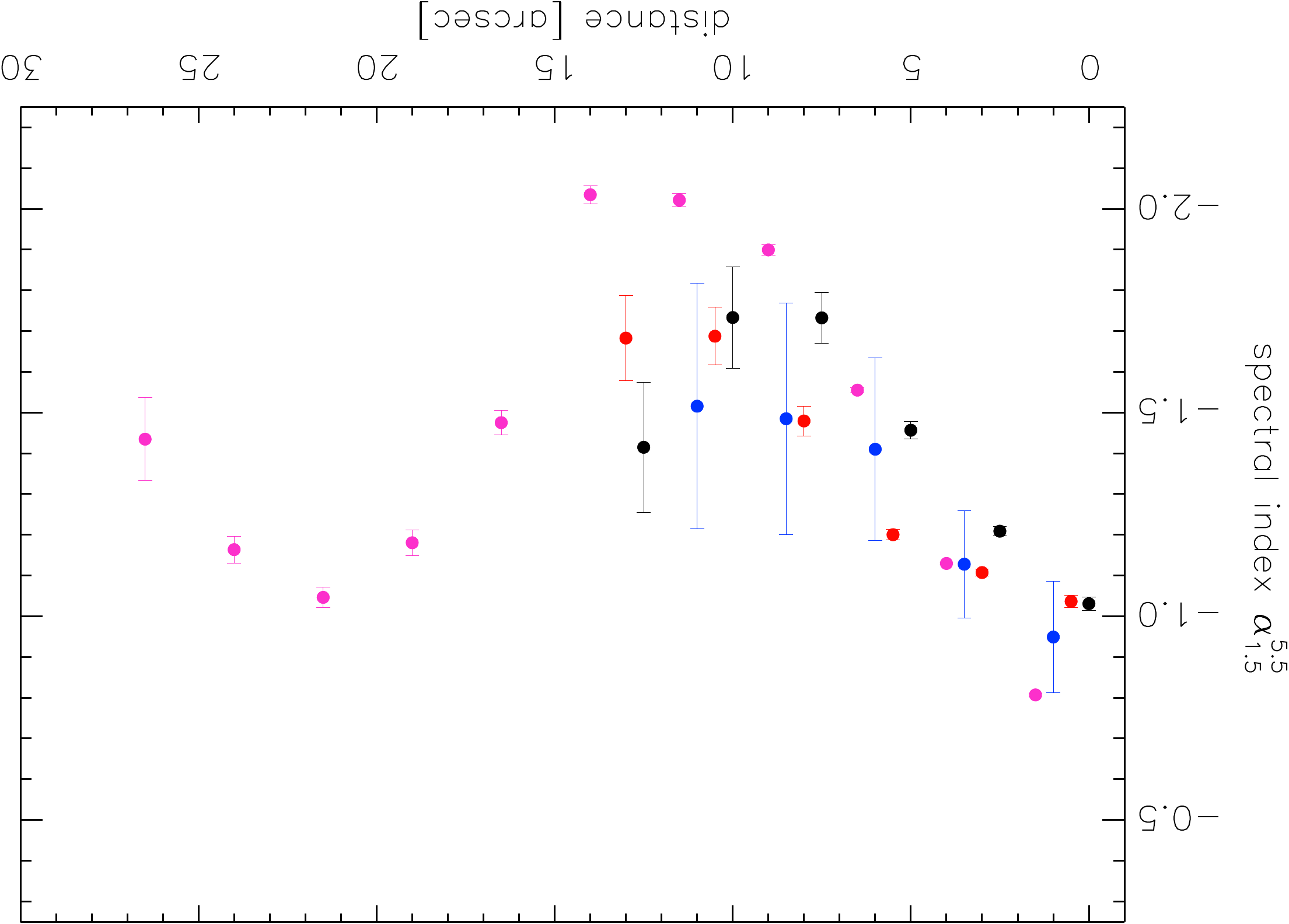}
\end{center}
\caption{Left: Regions where spectral indices (shown in the right panel) were extracted. The regions have a width of 2.5\arcsec. The region's colors are matched to the colored data points in the right panel. Right: Computed spectral indices between 5.5 and 1.5~GHz in the various regions indicated in the left panel (note that the 3~GHz flux densities were not used to compute the spectral indices in this figure so that we simply have a single spectral index value between the two most extreme frequency points). We used the maps at 2.5\arcsec~resolution (which were also used to compute the spectral index maps). The distance is increasing from east to west (left to right). The errors shown on this plot only include the statistical uncertainties due to the image noise (a systematic uncertainty would affect all plotted points in the same way). The x-axis values are offset to aid the visibility.}
\label{fig:spixregions}
\end{figure*}

The central NAT source shows clear evidence of spectral steepening in the higher resolution spectral index maps, from about $-0.5$ to $-2.5$ towards the NW. The steepening trend is expected for spectral ageing along the tails of the source.  The extracted spectra, from the maps at 2.5\arcsec~resolution, as a function of distance from the radio core are displayed in Figure~\ref{fig:spixregions}. In the lower resolution spectral index maps this
steepening is reduced, which is expected, since the reduced resolution causes mixing of emission from nearby regions (i.e., the relic) with flatter spectra. Evidence for spectral steepening along the tails is also found at the far SE tailed source ($z=0.5399$), from $-0.6$ at the core to $-1.6$ at the end of the tail.

The lobes of the foreground ($z=0.1546$) FRI source have a relatively flat spectral index of about $-0.5$ to $-0.6$ (within a distance of $\sim 0.5\arcmin$ ($\sim 80$~kpc) from the core), while the core has an inverted spectrum with $\alpha \approx +0.5$. Little steepening is seen along the lobes of the source, from about $-0.6$ to $-0.8$. The LAS of 3.5\arcmin~corresponds to a physical size of 560~kpc at the source  redshift.

Radio relics often display spectral index gradients, with the spectral index steepening in the direction of the cluster center \citep[e.g.,][]{2006AJ....131.2900C,2008A&A...486..347G,2010Sci...330..347V,2012MNRAS.426...40B,2012MNRAS.426.1204K, 2013A&A...555A.110S}. This spectral steepening is explained by synchrotron and Inverse Compton (IC) losses in the post-shock region of an outward traveling shock front. We also find a spectral index gradient in MACS~J0717.5+3745, with the spectral index decreasing from about $-0.9$ to  $\lesssim -1.6$ towards to west. This trend is particularly pronounced for R4 (lower left panel of Figure~\ref{fig:spix}).

This spectral steepening is also clearly seen in the spectral index profile (between 1.5 and 5.5~GHz) extracted across the relic in two regions, see Figure~\ref{fig:spixregions}. Hints of this  east-west trend of steepening across the relic were noted before by \citet{2009A&A...505..991V}. For both regions marked by the black and red boxes, we find steepening from $-1.0$ to about $-1.6$. No clear spectral index trends were reported by \citet{2009A&A...503..707B}, but this can be explained by the lack of signal-to-noise compared to our new \jvla\ observations.

The spectral index distribution for the radio relic around the central NAT source is more complex.
This is not too surprising given that the relic in MACS~J0717.5+3745 has an  irregular asymmetric morphology, likely the result of the complex quadruple merger event, and is projected relatively close to the cluster center, implying that the structure is not necessarily observed close to edge-on \citep{2012MNRAS.421.1868V}.

We also find evidence for EW spectral steepening, from about $-1.0$ to $-1.5$, across the brighter northern filament (blue region, Figure~\ref{fig:spixregions}).  However, the uncertainties are significant as is indicated on the plot.
The bright filament just north of R4 has a flatter spectrum ($-1.1$) than the halo emission in the vicinity. This is also the case for the filament at the southernmost part of the radio halo. The radio halo spectral index varies between $-1.2$ and about $\approx -2$ (the region south of R4).

\section{X-ray Results}
\label{sec:xrayresults}
\subsection{Global X-ray Properties}

\begin{figure*}
\centering
   \includegraphics[width=0.75\textwidth]{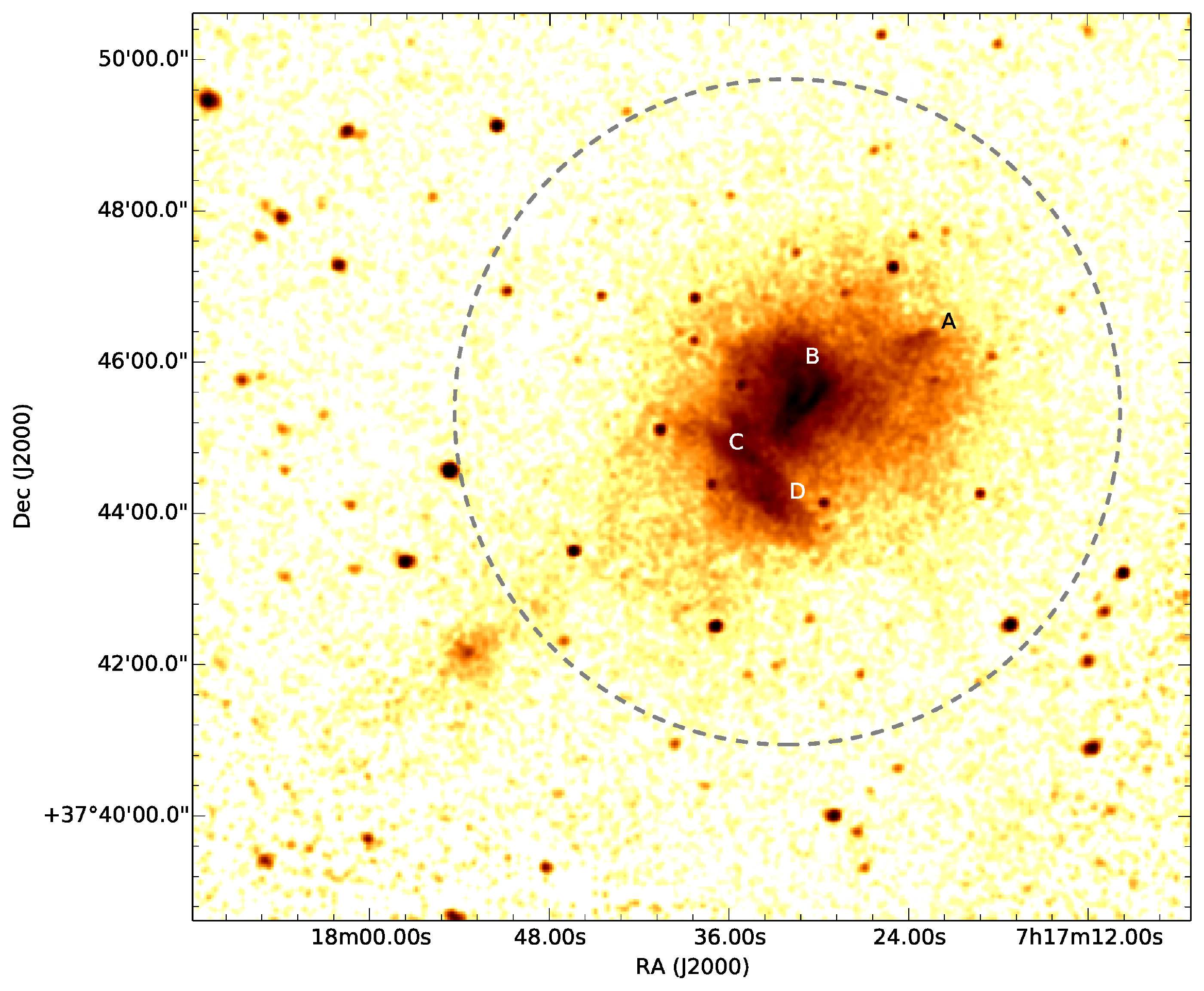}
   \caption{\chandra\ $0.5-4$~keV surface brightness map of MACS~J0717.5+3745. The image was vignetting- and exposure-corrected, and smoothed with a Gaussian of width 2\arcsec. The dashed-line circle shows  $R_{\rm 500}$ for the cluster. \label{fig:sxmap}}
\end{figure*}

\begin{figure*}[h]
\begin{center}
\includegraphics[angle =180, trim =0cm 0cm 0cm 0cm,width=1\columnwidth]{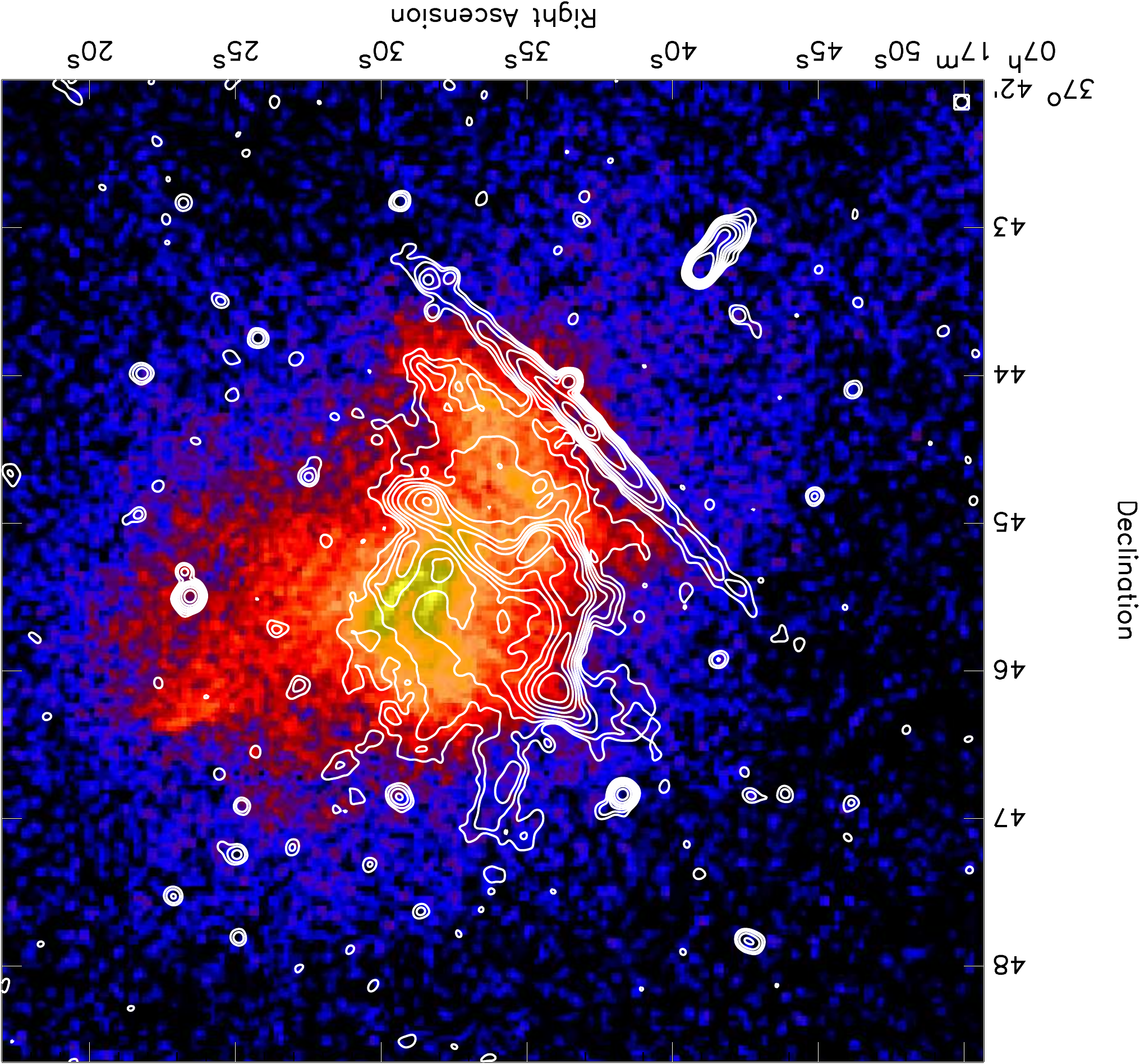}
\includegraphics[angle =180, trim =0cm 0cm 0cm 0cm,width=1\columnwidth]{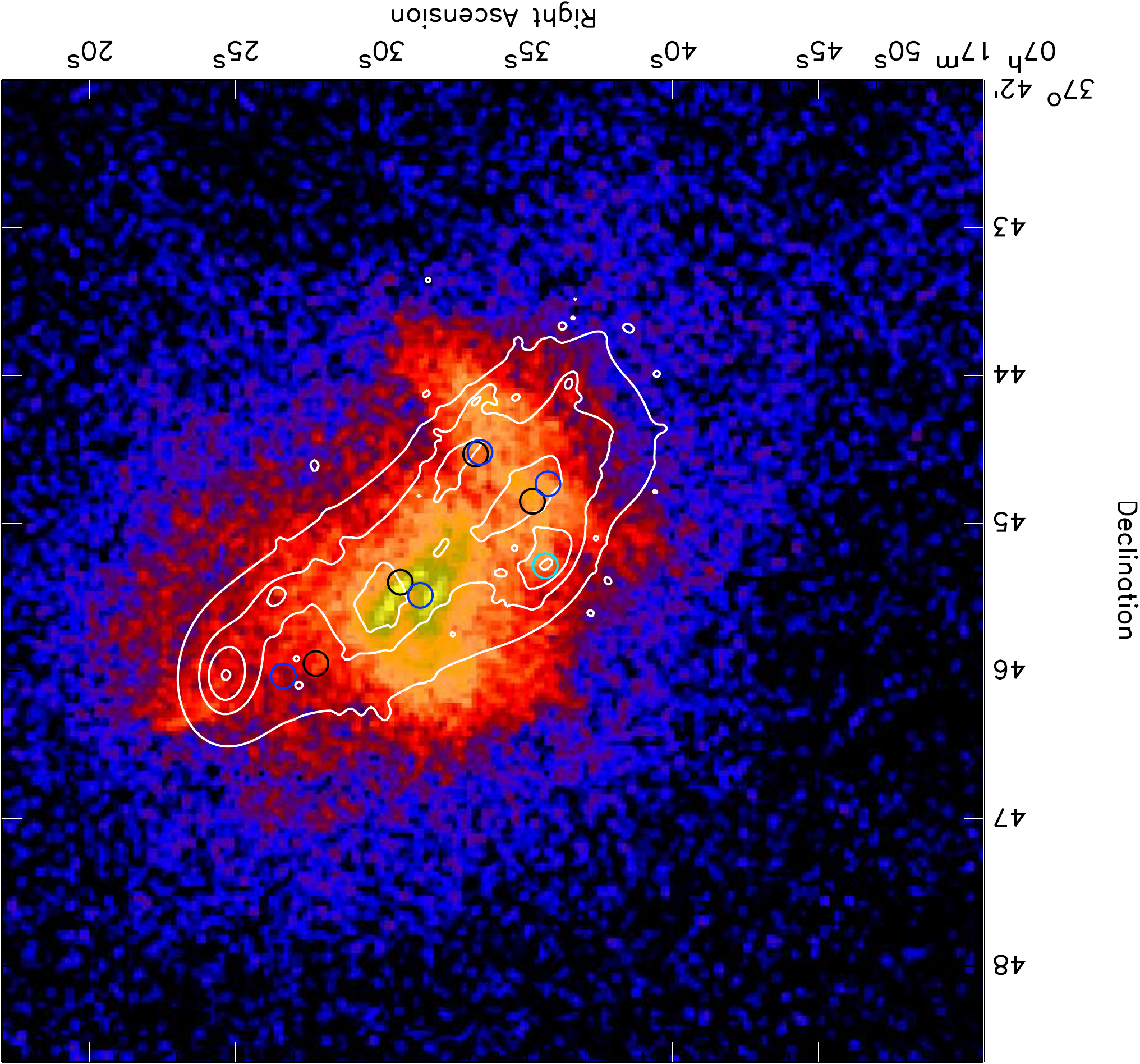}
\end{center}
\caption{Left: \chandra\ 0.5--4.0~keV X-ray image. Compact sources were removed and gaps were replaced by the average surface brightness in their surroundings (with Poisson noise added). Radio contours are from the 5\arcsec~resolution image and drawn at $[1,2,4,\ldots] \times 4\sigma_{\rm{rms}}$. Right: Same image as in the left panel but with the convergence map $\kappa = \frac{\Sigma}{\Sigma_{\rm{cr}}}$ (with $\Sigma_{(\rm{cr})}$ the (critical) mass surface density density) overlaid from \citet{2015ApJ...799...12I}. Contour levels are drawn at $\kappa = [1,1.5,2,4]\times0.8$. The positions of several mass components from \citet{2014ApJ...797...48J} and \citet{2016A&A...588A..99L} are indicated with black and blue circles, respectively. The cyan circle corresponds to an individual (massive) cluster galaxy \citep{2014ApJ...797...48J}.}
\label{fig:chandraradio}
\end{figure*}

\begin{figure*}[h]
\begin{center}
\includegraphics[angle =180, trim =0cm 0cm 0cm 0cm,width=0.9\textwidth]{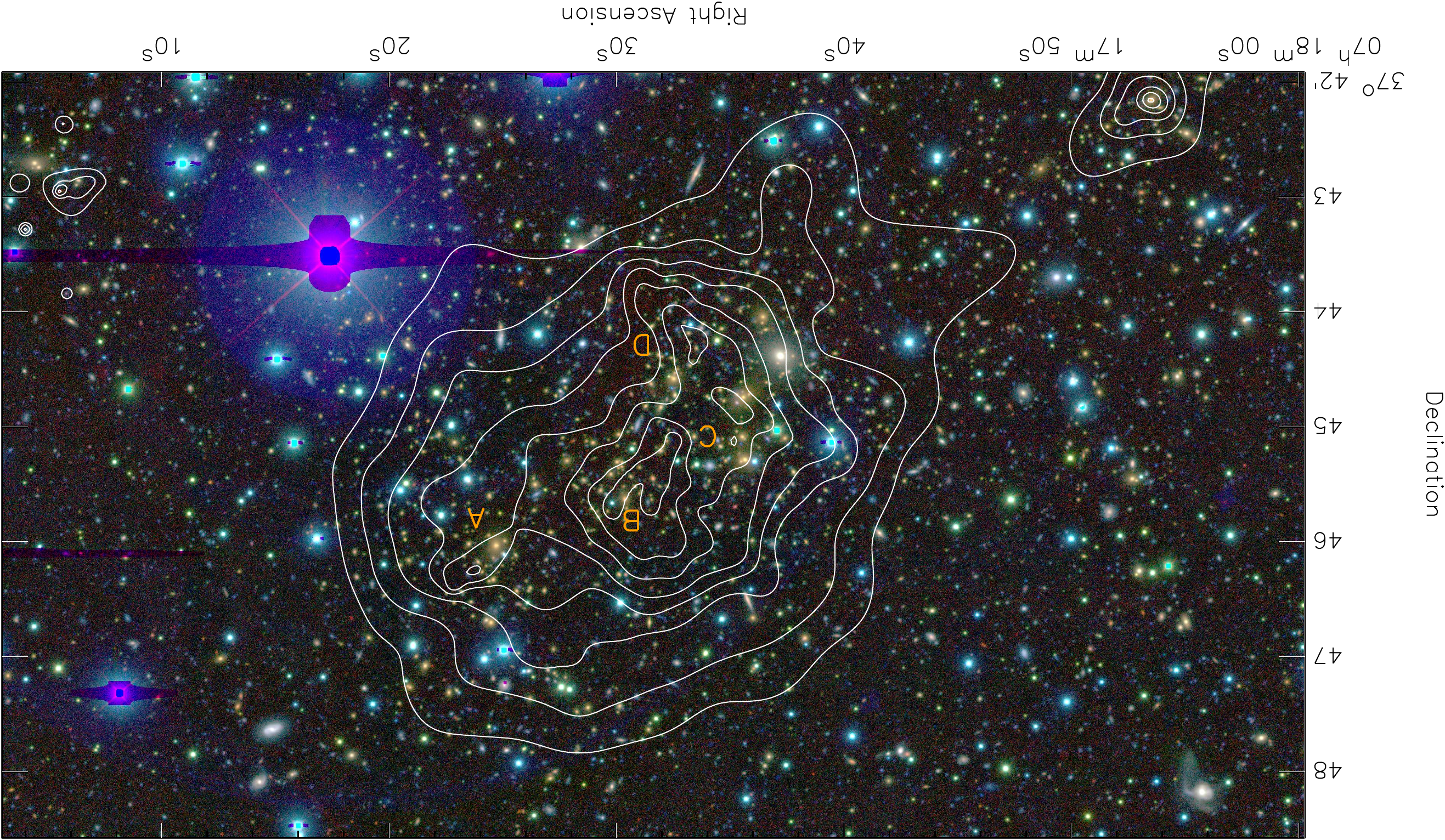}
\end{center}
\caption{Subaru B, I, and  CFHT Ks band color image of MACS~J0717.5+3745 \citep{2013ApJ...777...43M,2014ApJ...795..163U}. \chandra\ 0.5--4.0~keV contours, adaptively smoothed \citep{2006MNRAS.368...65E},  from Figure~\ref{fig:chandraradio} are overlaid. The contour levels are spaced according to $\propto (n)^{4/3}$, with $n=[1,2,3,\ldots$]. The subclusters A--D are labeled as in \citet{2009ApJ...693L..56M}.}
\label{fig:chandraoptical}
\end{figure*}

In Figure~\ref{fig:sxmap}, we present the  \chandra\ $0.5-4$~keV image of the cluster, vignetting- and exposure-corrected.
This image shows the main structures, as found earlier by \citet{2009ApJ...693L..56M}. The properties of the Mpc-scale X-ray filament to the southeast of the cluster will be discussed by \citet{gogreanfilament}.  \chandra\ images with \jvla\ radio contours and the surface mass density \citep[derived from a lensing analysis,][]{2015ApJ...799...12I} overlaid, are shown in Figure~\ref{fig:chandraradio}. An optical Subaru-CHFT image overlaid with X-ray contours is shown in Figure~\ref{fig:chandraoptical}. Four different substructures (A--D) are labeled, following \citet{2009ApJ...693L..56M}.

\begin{figure*}
\centering
  \includegraphics[ trim =0cm 2cm 0cm 0cm, angle=0,width=1.2\columnwidth]{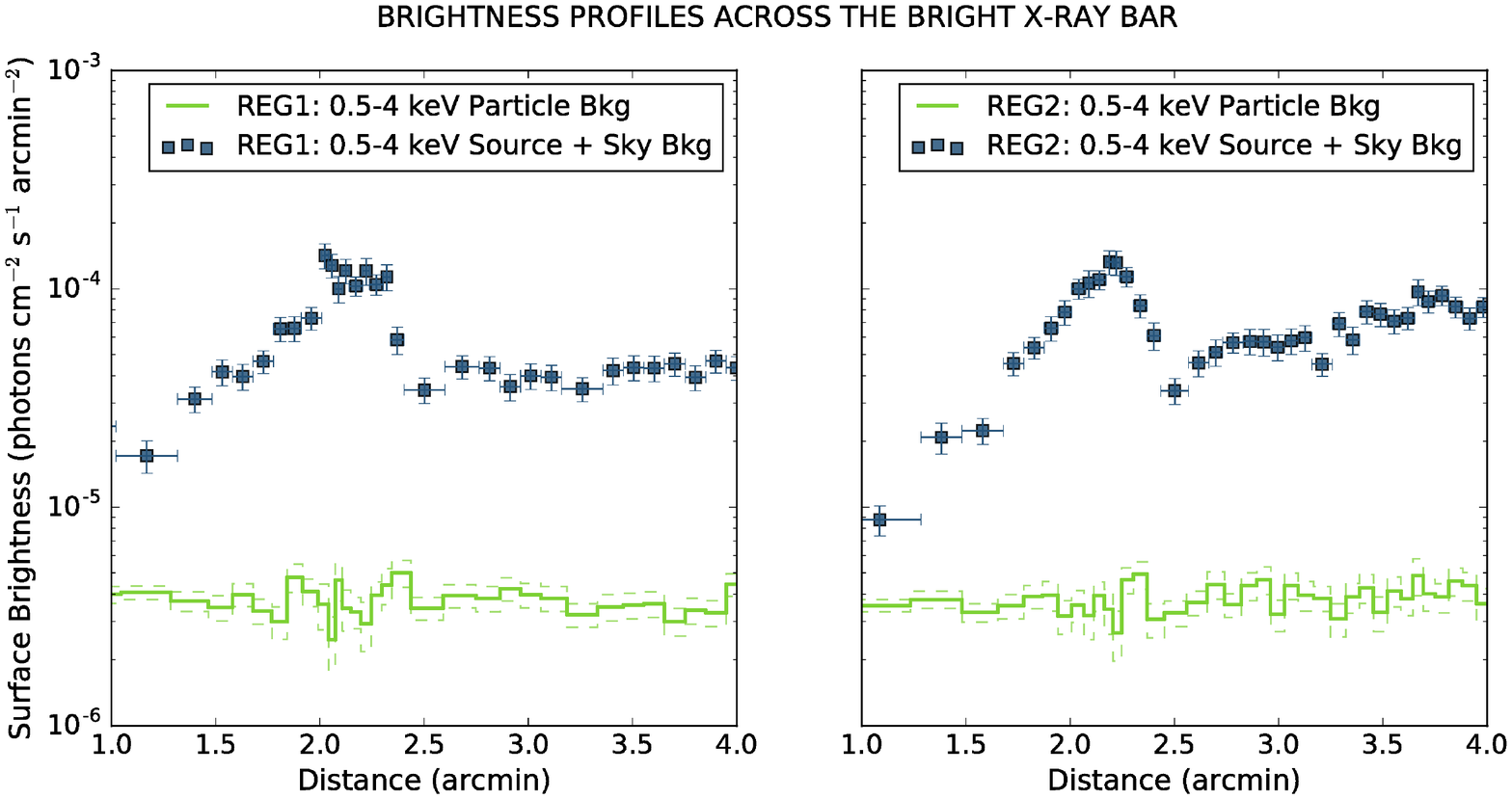}
  \includegraphics[angle=0,width=0.9\columnwidth]{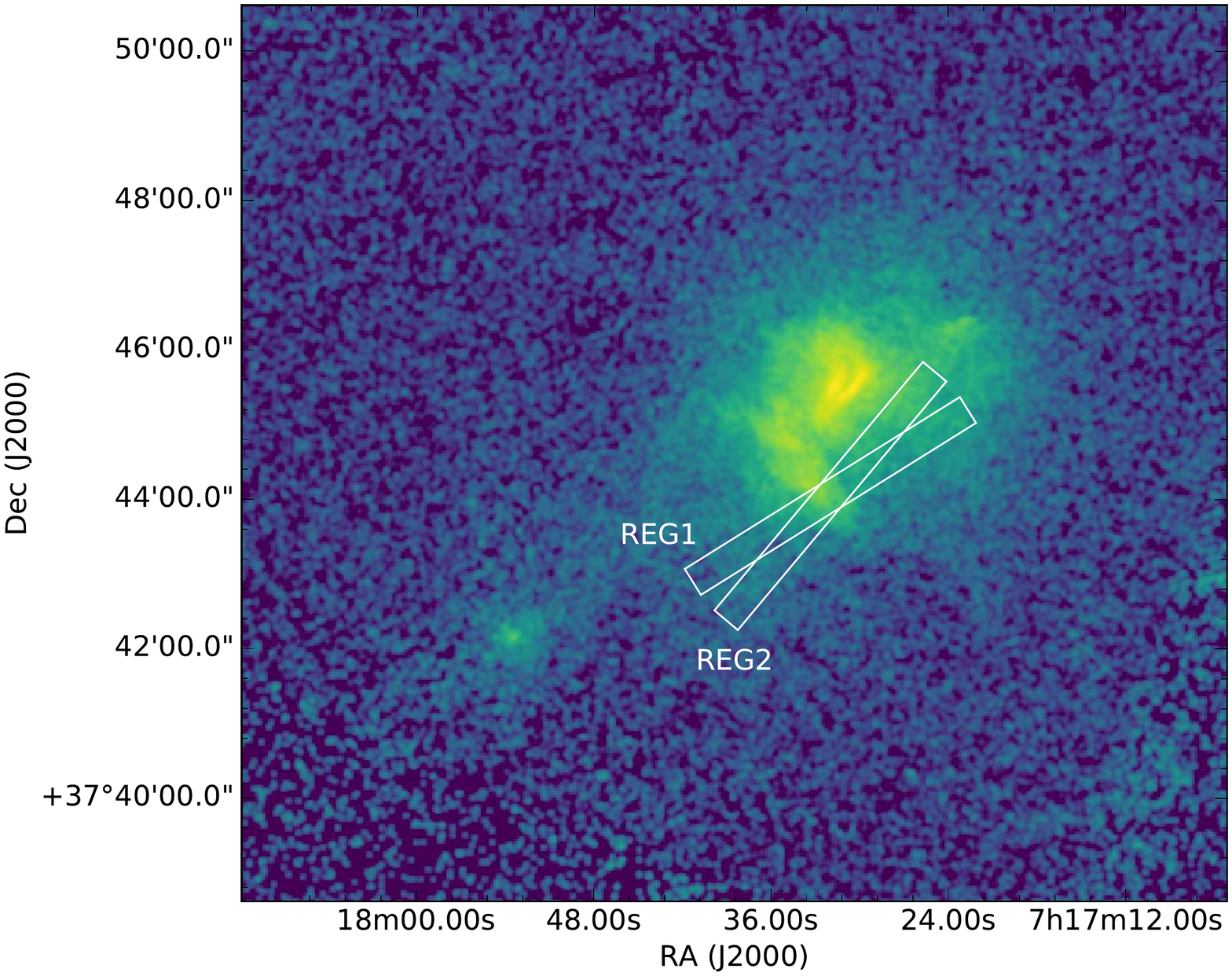}\vspace{0mm}
  \caption{X-ray surface brightness profiles across the bar (SE to NW) in two regions as indicated in the right panel. The bar shows a hint of an edge on its western side, located at a distance of about 2.4\arcmin. The instrumental background is shown in green, with the uncertainty ranges on the background shown in dashed green lines.}
\label{fig:barprofile}
\end{figure*}

The X-ray emission of the cluster is complex, consisting of a bar-shaped structure to the southeast with a size of $800\times300$~kpc.  The bar consists of two separate components (C and D, see Figures~\ref{fig:sxmap} and~\ref{fig:chandraoptical}). These two components are likely associated with two separate merging subclusters and are also detected in the mass surface density map. X-ray surface brightness profiles across the bar along two rectangular boxes is presented in Figure~\ref{fig:barprofile}. These profiles show that the western edge of the bar is cut off more abruptly than the eastern edge. We did not attempt to fit a density model to the edge because of the unknown (and likely complex) geometry.

The brightest part of the ICM consists of a V-shape structure, which is associated with major mass component~B. To the northwest, an elongated, bullet-like X-ray substructure is seen, with a sharp boundary on its northern edge. This structure seems to be associated with mass component~A, and is also seen in the mass surface density map from \citet{2015ApJ...799...12I}.  However, \citet{2014ApJ...797...48J} and \citet{2016A&A...588A..99L} place the center of the westernmost mass component about 0.5\arcmin~east of the center of the X-ray component. We discuss this ``fly-through'' bullet-like core in more detail in Section~\ref{sec:fly-throughcore}. A small ``clump'' of gas is  found just north of the bar (again best seen in Figure~\ref{fig:sxmap}, located at the cyan circle in Figure~\ref{fig:chandraradio}).

From Figure~\ref{fig:chandraradio} we find that the radio filament north of R4 is  aligned with the SW part of the V-shaped structure.
The southernmost radio filament (Figure~\ref{fig:3res}) coincides with the southern end of the X-ray bar. The two northern filaments (north of R1) are located in the faint X-ray outskirts of the cluster.

In the SE, the radio halo emission roughly follows the outline of the bar. North of R4 the halos follows the bright X-ray region consisting of the V-shaped structure and emission north of it. The western part of the cluster is devoid of  diffuse radio emission.


To measure the global X-ray properties of the cluster, we extracted \chandra\ spectra in a circle with a radius of $R_{\rm 500}=1.69$~Mpc \citep{2010MNRAS.406.1773M} around ${\rm RA} = 07^{\rm{h}}17^{\rm{m}}32\fs1$ and ${\rm DEC} = +37\degr45\arcmin21\arcsec$. The spectra were instrumental background-subtracted, and modeled as the sum of absorbed thermal ICM emission and sky background emission. The sky background model was fixed to the model summarized in Table~\ref{tab:skybkg}. The temperature, metallicity, and normalization of the thermal component describing ICM emission were left free in the fit. 

We measured $T_{\rm 500} = 12.2_{-0.4}^{+0.4}$~keV, $Z=0.21\pm 0.03\;Z_{\sun}$, and a $0.1-2.4$~keV luminosity of $(2.35\pm 0.01)\times 10^{45}$~erg~s$^{-1}$.

\subsection{Temperature Map}

\begin{figure*}
  \includegraphics[angle=0,width=\columnwidth]{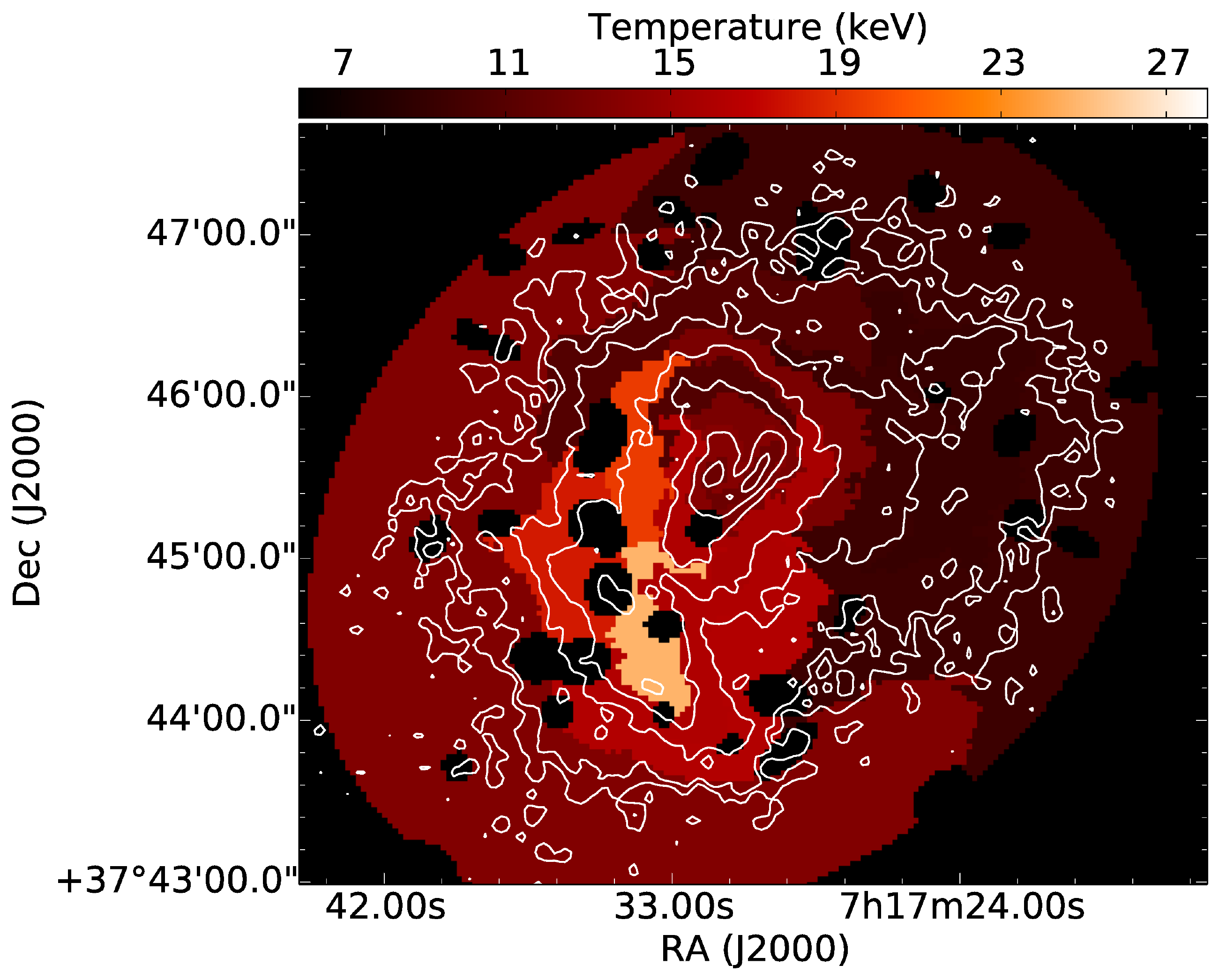}\hfill
  \includegraphics[angle=0,width=\columnwidth]{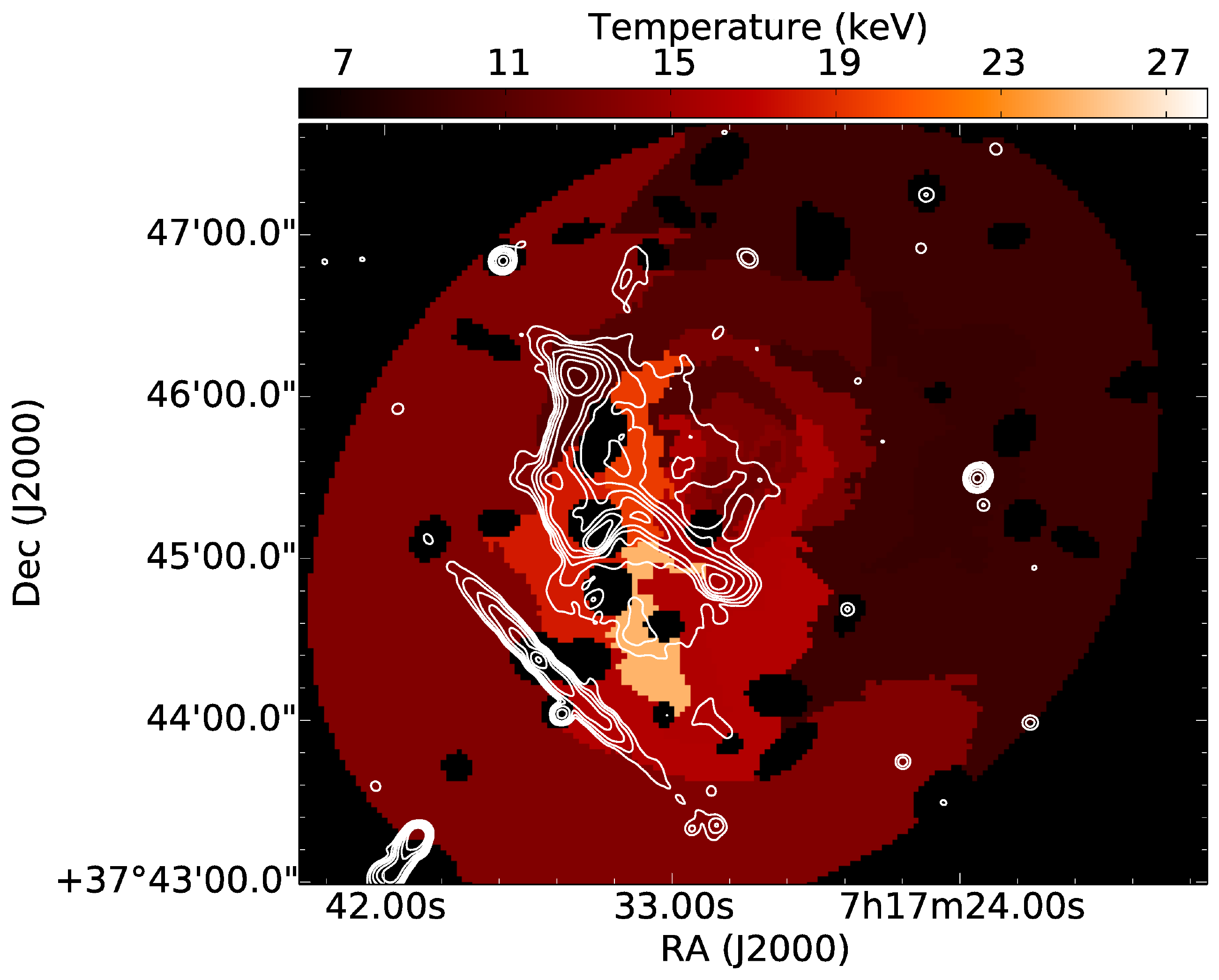}
  \caption{Temperature map of MACS~J0717.5+3745. Overlaid are \chandra\ $0.5-4$~keV surface brightness contours (left; based on the image in Figure~\ref{fig:sxmap}) and \jvla\ radio contours (right; from Figure~\ref{fig:3res} middle panel). The X-ray contours are drawn at $[0.013,0.026,0.052,0.104,0.208,0.416,0.832,1.664]\times 10^{-6}$~photons~cm$^{-2}$~s$^{-1}$. The radio contours are drawn at $[1,2,3\ldots]\times 4\sigma_{\rm{rms}}$.}
\label{fig:temp+contours}
\end{figure*}

To map the ICM temperature, we used {\tt CONTBIN} \citep{Sanders2006} to bin the surface brightness map smoothed to a ``signal''-to-noise of 10 in individual regions with a uniform ``signal''-to-noise ratio of 55. Here, by ``signal'' we refer not only to the ICM signal, but rather to ICM and sky background signal combined; the noise is the instrumental background emission. We extracted total spectra and instrumental background spectra from each of the individual regions, and modeled them as the sum of absorbed thermal emission from the ICM and sky background emission. The parameters of the sky background model were fixed to the values in Table~\ref{tab:skybkg}. The ICM metallicity was fixed to $0.21$~$Z_{\sun}$. Figure~\ref{fig:temp+contours} shows the resulting temperature map. An interactive version of the map, which includes uncertainties on the best-fitting spectral parameters at the $90\%$ confidence level, is available at \url{https://goo.gl/KtE33D}. 

In Figure~\ref{fig:temp+contours}, we show the temperature map with overlaid X-ray and radio contours. \citet{2009ApJ...693L..56M} argued that the V-shaped region (subcluster~B) contains a cool core remnant with a temperature of $\sim 5$~keV. However, we find no evidence of such low-temperature gas, instead measuring a temperature of $\sim 12$~keV in the V-shaped region. The results reported by \citet{2009ApJ...693L..56M} were based only on ObsID 4200. Neither using only ObsID 4200, nor changing the region used to measure the temperature allowed us to obtain a temperature lower than 8 keV (with the $90\%$ confidence level uncertainties considered). We also did a separate analysis that followed that of \citet{2009ApJ...693L..56M} more closely: we used blank-sky event files, fitted the ICM with a MEKAL model, fixed the abundance to $0.3$ solar, and fixed the absorption to $7.11\times 10^{20}$~cm$^{-2}$. Again, the temperature we obtained was above 9~keV at the $90\%$ confidence level.

Our temperature map reveals an extremely hot region in the SSE part of the cluster center, with a temperature $\gtrsim 20$~keV. This hot region is associated with the bar-shaped region of enhanced surface brightness seen in Figure~\ref{fig:sxmap}. \citet{2009ApJ...693L..56M} reported another possible cool core remnant in the W part of this region, where they measured a temperature of $8.4\pm 3.6$~keV ($1\sigma$ uncertainties, region A22 in their publication). This temperature was significantly lower than the temperatures reported in adjacent regions, which all had $>15$~keV gas. Choosing a region that approximates that of \citet{2009ApJ...693L..56M}, we measure $13.8_{-3.0}^{+4.1}$~keV ($1\sigma$ uncertainties). While this temperature is consistent with that measured by \citet{2009ApJ...693L..56M}, it is also consistent with the temperatures of the adjacent regions. 


In conclusion, we find temperatures above $\sim 10$~keV throughout the ICM, with a temperature peak of $>20$~keV in the X-ray bright, bar-shaped region SSE of the radio relic. Similarly, \citet{2012ApJ...761...47M} did also  not report temperatures below $\sim10$~keV using XMM-Newton and \chandra\ observations of the cluster. Therefore, we do not confirm the temperatures of the cool regions reported by \citet{2009ApJ...693L..56M}. The V-shaped region does seem to be cooler than its immediate surroundings, but not at the level as reported by \citet{2009ApJ...693L..56M}.

\subsection{Fly-Through Core}
\label{sec:fly-throughcore}

\begin{figure}
   \centering
    \includegraphics[trim =0cm 0cm 0cm 0cm, width=1.0\columnwidth]{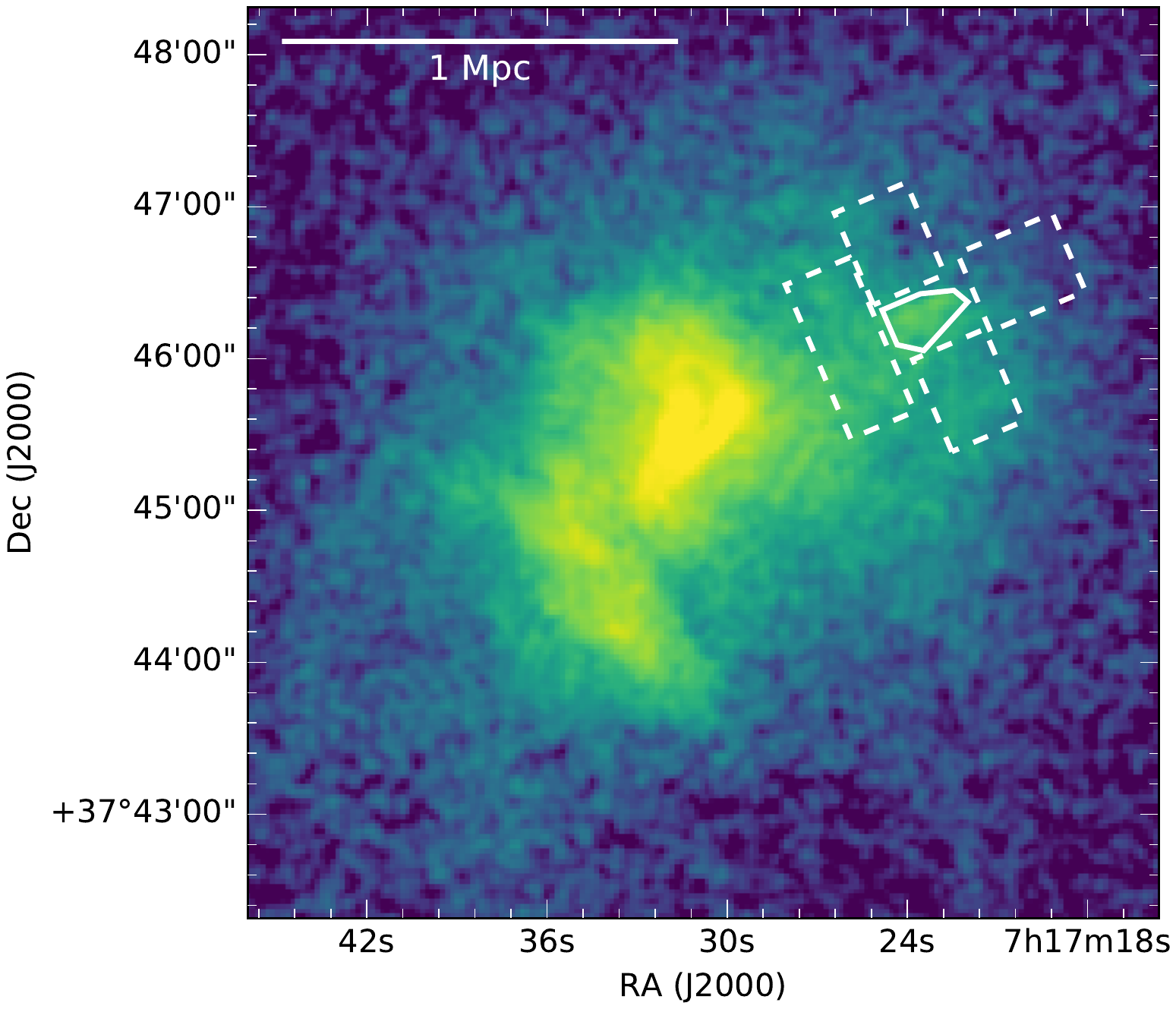}\vspace{-30mm}\hfill
    \caption{Regions used in the spectral analysis. The regions of main interest are drawn in solid lines, while the regions used to characterize the contaminating/surrounding emission are drawn in dashed lines. The best-fitting parameters obtained for the gas in these regions are listed in Table~\ref{tab:spectra}. \label{fig:fil}}
\end{figure}

\begin{figure*}
    \includegraphics[width=\columnwidth]{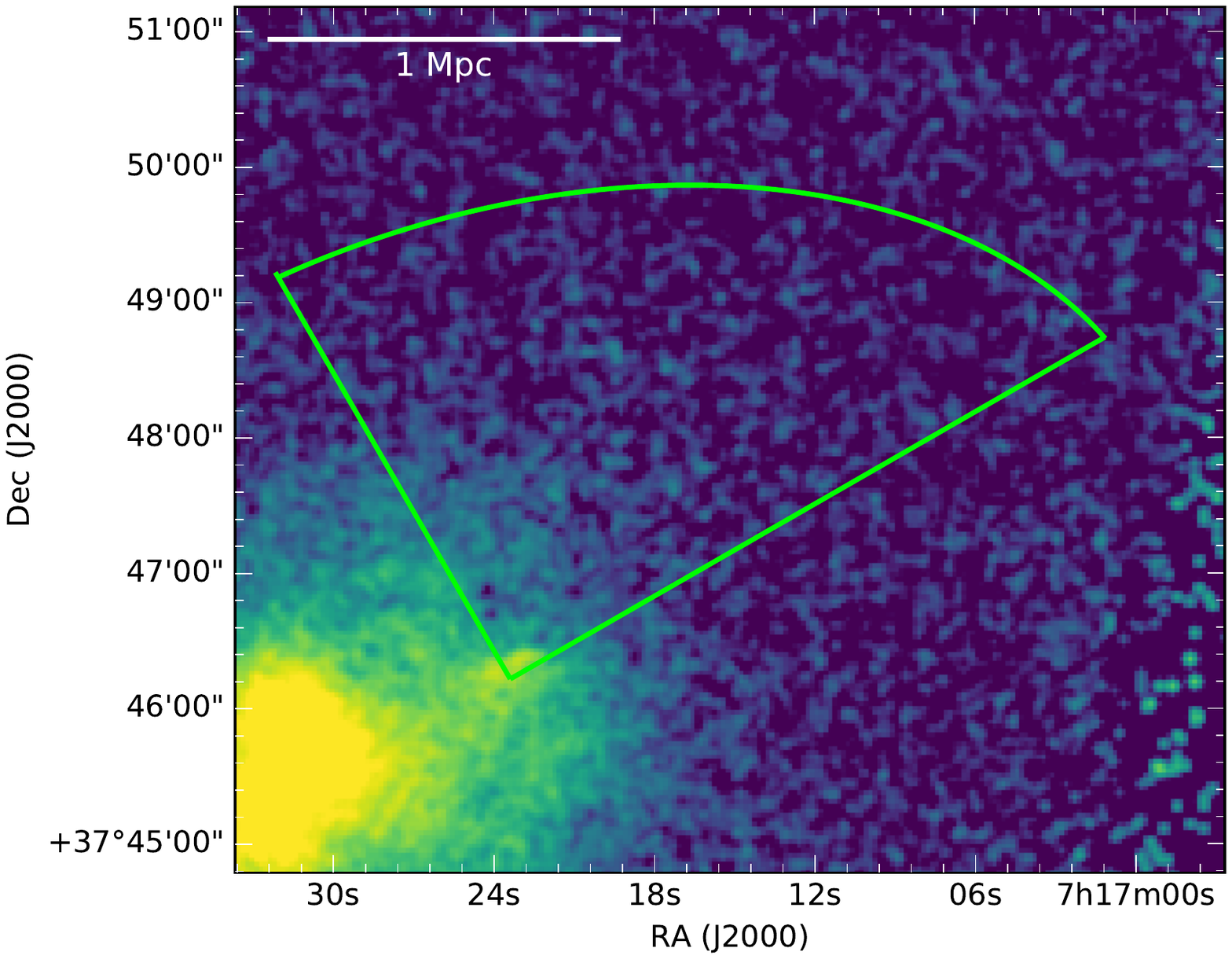}
    \includegraphics[width=0.5\textwidth]{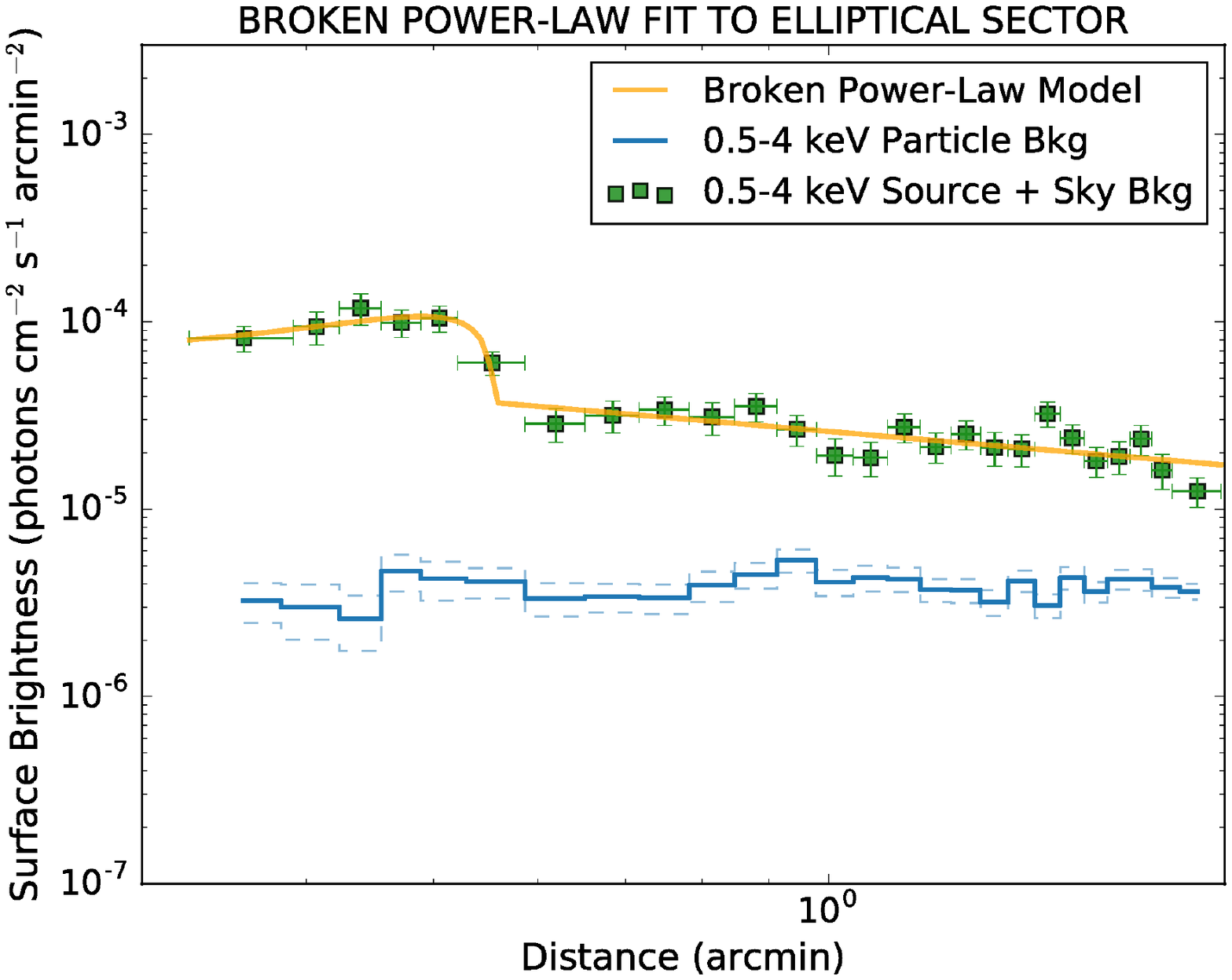}
    \vspace{-30mm}
    \caption{\emph{Left:} Sector used to model the surface brightness profile in front of the core [GO: needs to be updated]. \emph{Right:} Surface brightness profiles and best-fitting model. The instrumental background is shown in blue, with the uncertainty ranges on the background shown in dashed blue lines. For the elliptical sector, the surface brightness is plotted against the major axis of the ellipse. The best-fitting parameters of the broken power-law model are listed in Table~\ref{tab:bknpow}. \label{fig:sectors}}
\end{figure*}

\begin{table*}
  \caption{Best-fitting parameters of the broken power-law model fitted to the surface brightness of the core. Uncertainties are quoted at the $1\sigma$ level. The region from which the surface brightness profile was extracted, as well as the profile and best-fitting model, are shown in Figure~\ref{fig:sectors}. \label{tab:bknpow}}
    \begin{threeparttable}
      \begin{tabular}{c c c c c c}
      \hline\hline
              $\alpha_1$\tnote{a} & $\alpha_2$\tnote{b} & Normalization & $r_{\rm break}$ & $n_1/n_2$\tnote{c} &  Sky background \\
                               &        &     [photons cm$^{-2}$ s$^{-1}$ arcmin$^{-2}$]       & [arcmin]   & &    [photons cm$^{-2}$ s$^{-1}$ arcmin$^{-2}$] \\
              \hline
              $-0.45^{+1.3}_{-1.5}$ & $1.05^{+0.11}_{-0.10}$ & $6.34^{+1.47}_{-1.18}  \times 10^{-4}$ & $0.56^{+0.01}_{-0.01}$ & $3.27^{+0.44}_{-0.46}$ & $(1.30 \pm 0.14) \times 10^{-6} $ \\
              \hline \hline
      \end{tabular}
      \begin{tablenotes}
              \item[a] Power-law index at $r < r_{\rm break}$.
              \item[b] Power-law index at $r > r_{\rm break}$.
              \item[c] Density jump across the discontinuity.
      \end{tablenotes}
    \end{threeparttable}
\end{table*}

Approximately 0.7~Mpc NW from the cluster center, there is a X-ray core (Figure~\ref{fig:fil}) with a tail extending $\sim 200$~kpc towards the SE, roughly in the direction of the large-scale galaxy filament in the SE. This morphology suggests that this core, seen ``flying'' through the ICM of MACS~J0717.5+3745 and ram-pressured stripped by the cluster's dense ICM, traveled NW along the SE filament and is seen after it traversed the brightest ICM regions. 
In essence, the core is analogous to a later stage of the group currently seen within the filament.

The core is embedded (at least in projection) in the ICM of MACS~J0717.5+3745. To determine the core's physical properties, we  modeled the contamination from the ICM by extracting spectra N and S of the core. These spectra were modeled with a thermal component with a metallicity of 0.2~solar. We assumed the spectral properties were the same in the N and S regions. The spectra of the core were modeled as the sum of emission from the contaminating ICM and from the core itself. The spectra of the core and of the regions N and S of it were modeled in parallel. The best-fitting results are summarized in Table~\ref{tab:spectra} and the regions are indicated on Figure~\ref{fig:fil}. The temperature of the core, $6.82_{-1.36}^{+1.88}$~keV, is consistent with the temperatures N and S of the core, in regions that are approximately at the same distance from the cluster center as the core. We also compared the core temperature with the temperatures ahead of (NW) and behind (SE) the core. The temperature decreases from $10.89_{-1.27}^{+2.05}$~keV behind the core, to $5.06_{-0.98}^{+1.61}$~keV ahead of the core. From these temperature measurements, we therefore find no evidence of a core colder than its surroundings, nor of a temperature discontinuity (either a shock or a cold front) ahead of the core.

\begin{table}
  \caption{Parameters of the regions used for the spectral analysis of the fly-through core. The regions are shown in Figure~\ref{fig:fil}. Uncertainties are quoted at $1\sigma$ level. \label{tab:spectra}}
    \begin{threeparttable}
      \begin{tabular}{l c c}
              \multicolumn{3}{c}{{ }} \\
              \hline\hline
                Model Component & Temperature\tnote{a} & Normalization\tnote{b} \\
              \hline
               Core & $6.82_{-1.36}^{+1.88}$ & $3.41_{-0.25}^{+0.29} \times 10^{-4}$ \\
               N+S of Core & $7.47_{-0.86}^{+1.11}$ & $2.08_{-0.78}^{+0.77} \times 10^{-4}$  \\
               Ahead of Core & $5.06_{-0.98}^{+1.61}$ & $8.52_{-0.77}^{+0.88} \times 10^{-5}$  \\
               Behind Core & $10.89_{-1.27}^{+2.05}$ & $3.92_{-0.09}^{+0.10} \times 10^{-4}$  \\
               \hline
               \hline
      \end{tabular}
      \begin{tablenotes}
              \item[a] Units of keV.
              \item[b] Units of cm$^{-5}$~arcmin$^{-2}$ for the thermal components, and photons~keV$^{-1}$~cm$^{-2}$~s$^{-1}$~arcmin$^{-2}$ at 1~keV for the power-law components.
      \end{tablenotes}
    \end{threeparttable}
\end{table}

A cold front and a shock front would be expected ahead of the core, similarly to the features seen in the Bullet Cluster \citep{2002ApJ...567L..27M} and in front of the group NGC 4839 infalling into the Coma Cluster \citep{Neumann2001}, see also the review by \cite{2007PhR...443....1M}. 
We searched for possible evidence of a cold/shock front by modeling the surface brightness profile of the group. 
The sector from which the surface brightness profiles was extracted is shown in Figure~\ref{fig:sectors} (left panel). We chose an elliptical sector with an opening angle and ellipticity aligned with a possible edge observed by eye in the surface brightness map. The model fitted to the surface brightness profile is shown in the right panel of Figure~\ref{fig:sectors}. The surface brightness profile extracted from the circular sector is well-fitted by a broken power-law density model. In this profile, there is an edge near $\sim 0.5\arcmin$ -- $0.6\arcmin$. The best-fitting model has a density jump of $3.3\pm 0.4$ at $\approx 0.56\arcmin$ from the center of the sector. The best-fitting parameters for the broken power-law model are summarized in Table~\ref{tab:bknpow}.

The density discontinuity is at the very edge of the core. Therefore, we speculate that the discontinuity is associated with a cold front rather than with a shock front. The failure to find a temperature discontinuity associated with the density jump is likely due to poor count statistics and emission from hot gas
projected onto the core.  The latter also dilutes the observed density jump, in which case our measurement of the jump amplitude is only a lower limit\footnote{This applies to the situation were the emission from the hot gas exceeds the emission from outside the jump
in the broken power-law model}.

\section{Discussion}
\label{sec:discussion}

\subsection{Origin of the radio relic}

Radio relics are thought to trace relativistic electrons that are accelerated or re-accelerated at shocks. The presence of a powerful radio relic in the cluster MACS~J0717.5+3745 is therefore consistent with the cluster undergoing a violent merger event. In fact, the \chandra\ temperature map indicates that the relic traces a hot shock-heated region with temperatures of $\sim 20$~keV and higher. If we interpret the observed spectral index trends across the relic, Figure~\ref{fig:spixregions}, as due to electrons cooling in the post-shock region, then the shock should be located at the eastern boundary of the relic and the post-shock region is located to the west of that.

\begin{figure}[h!]
   \includegraphics[angle=0,width=\columnwidth]{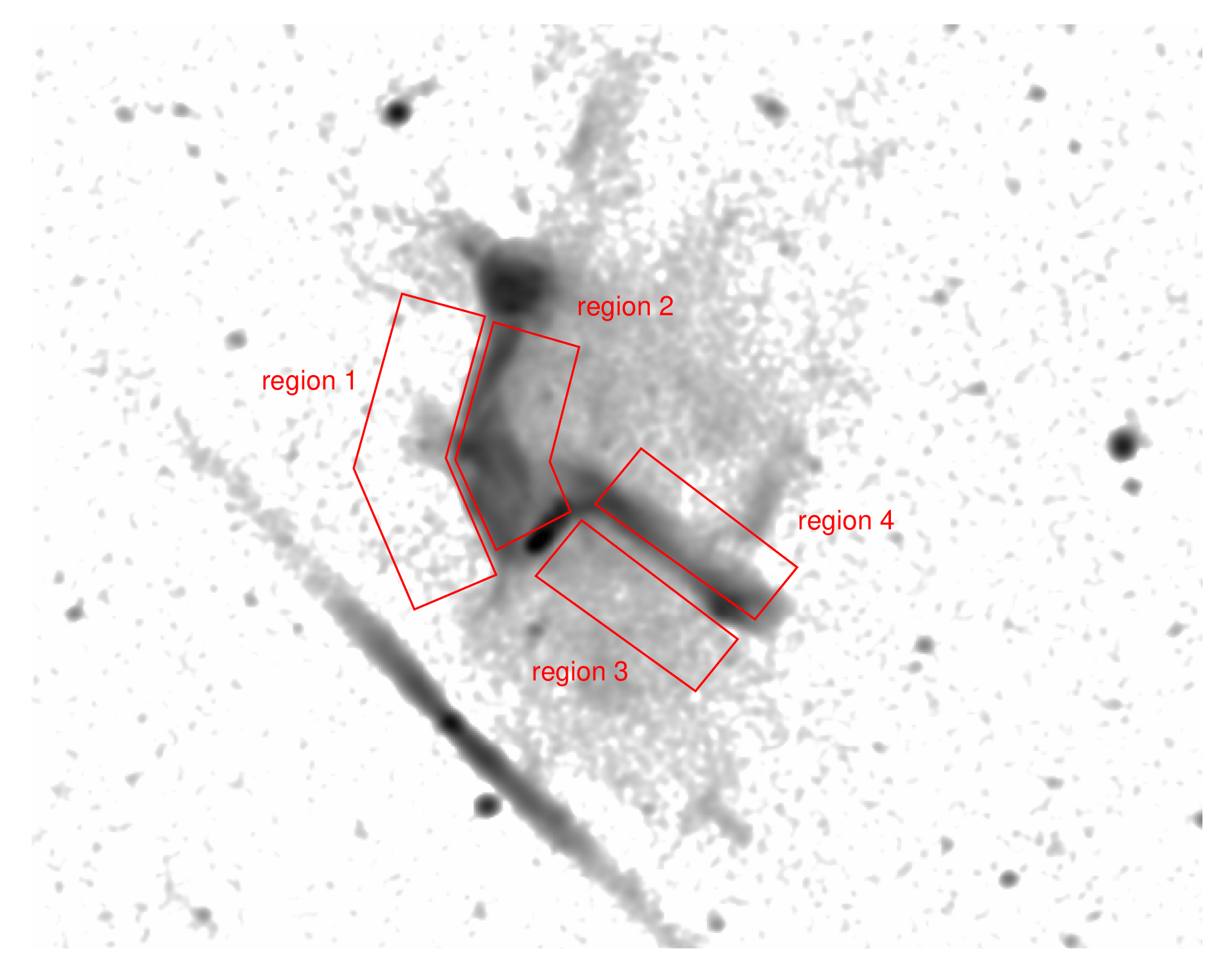}
   \caption{Regions where we extracted the temperatures on top of the wide-band 1.0--6.5~GHz radio image.}
   \label{fig:Tregion}
\end{figure}

We extracted temperatures on the eastern side of the relic ($T_1$, the putative pre-shock region; regions~1 and~3) and around the putative shock downstream region ($T_2$; regions~2 and ~4). The regions are indicated in Figure~\ref{fig:Tregion}. {For the northern part of the relic, we find $T_1=20.0^{+5.1}_{-3.7}$~keV and $T_2=20.3^{+12.6}_{-4.6}$~keV (regions 1, 2). For the southern part we measure $T_1=27.1^{+8.6}_{-5.6}$~keV and $T_2=16.6^{+3.1}_{-2.-}$~keV (regions 3, 4). So it is hard to say from the temperatures where the pre- and post-shock regions are. Since the relic is at least partly located in the cluster outskirts (the R1 and R2 part) and the X-ray emissivity is roughly proportional to the density squared the \chandra\ temperatures do not necessarily probe the actual pre- and post-shock gas but rather hot regions of higher density, with the relic projected close to it. This is particularly relevant for the southern part of the radio relic. We also do not detect any X-ray surface brightness edges associated with the relic. This might imply that the shock surface is not seen very close to edge-on and/or projection effects are important, or the Mach number is rather low.}

Therefore we conclude that given the complexity of the merger event and unknown projection effects, the precise relation between the relic and location of the hot gas remains uncertain.

\subsubsection{Acceleration mechanisms}

For relics, an important question is by which mechanism the synchrotron emitting electrons are accelerated. The standard scenario proposed by \citet{1998A&A...332..395E} is that particles are accelerated at shocks via the DSA mechanism. A problem with this scenario is that shocks in clusters generally have Mach number of $\mathcal{M} \lesssim 3$, and the acceleration of electrons from the thermal pool is thought to be very inefficient for these low Mach numbers, in apparent conflict with the presence of bright radio relics. 
In this case an unrealistic fraction of the energy flux through the shock surface \citep{2011ApJ...728...82M,2016MNRAS.461.1302E,2016ApJ...818..204V} needs to be converted to the non-thermal electron population.

The DSA mechanism should also accelerate protons to relativistic energies. These  protons then interact with the thermal gas to produce $\gamma$-rays. \citet{2014MNRAS.437.2291V,2015MNRAS.451.2198V,2016MNRAS.459...70V} show that the observed $\gamma$-rays upper limits are in tension with the relative acceleration efficiency of electrons and protons that is expected from DSA. Tension with DSA has also been found from the discrepancy between the measured Mach numbers from X-ray observations and  the radio spectral index (see Equation~\ref{eq:inj-mach}) for some relics \citep[e.g.,][]{2015PASJ...67..113I,2015A&A...582A..87A,2016ApJ...818..204V}.

PIC simulations  show that electrons can be accelerated from the thermal pool via the SDA mechanism, which would solve some of the problems with DSA \citep{2014ApJ...794..153G,2014ApJ...797...47G,2014ApJ...783...91C}. Another model to solve the low acceleration efficiency of standard DSA is that of re-acceleration of fossil electrons \citep[e.g.,][]{2005ApJ...627..733M,2008A&A...486..347G,2011ApJ...734...18K,2012ApJ...756...97K, 2013MNRAS.435.1061P}. 
These fossil electrons could, for example, originate from the (old) lobes of radio galaxies. Indeed  observations provide some support for this scenario because of the complex morphologies of some relics, suggesting a link with a nearby radio galaxy in a few select cases \citep{1991A&A...252..528G,2013ApJ...769..101V,2014ApJ...785....1B,2015MNRAS.449.1486S}. The most compelling case for re-acceleration has been found in the merging cluster Abell~3411-3412 \citep{a3411}. Here a tailed radio galaxy is seen connected to a relic. In addition spectral flattening is observed at the location where the fossil plasma meets the relic and at the same location an X-ray surface brightness edge is observed.

\subsubsection{Evidence for re-acceleration in  MACS~J0717.5+3745}
We argue that  the NAT galaxy in MACS~J0717.5+3745 provides another  compelling case for particle re-acceleration because (1)
the NAT galaxy is a spectroscopically confirmed cluster member,(2) we observe a morphological connection between the relic and NAT source, (3) there is evidence for hot shock-heated gas at the location of the radio relic (with the caveat of unknown projection effects), and (4) we can trace the spectral index across the tails of this galaxy until they fade into the relic. After fading into the relic the spectral index flattens again (Figure~\ref{fig:spixregions}, right panel magenta points), as is expected in the case of re-acceleration.

For a NAT source, we expect to start with a power-law radio spectrum, the radio spectrum then steepens  progressively along the tails of the NAT source due to synchrotron and IC losses. Apart from spectral steepening, the spectral curvature should also increase along the tails due to these energy losses. When the fossil electrons pass through the shock, they are re-accelerated and the spectral index flattens again. In MACS~J0717.5+3745 we observe this expected trend.

In the case of re-acceleration, the radio injection spectral index is set by the Mach number of the shock, unless the index of the fossil distribution is flatter than what would be produced by the re-acceleration process.  Following \citet{2005ApJ...627..733M},  we start with a power-law momentum fossil electron distribution
\begin{equation}
f_{\rm{fossil}} (p) \propto p^{-s_{\rm{fossil}}} \mbox{ ,}
\end{equation}
the distribution after re-acceleration (not considering energy losses) can be given by 
\begin{equation}
f_{\rm{inj,re}} (p) \propto p^{-s_{\rm{inj,re}}}  \mbox{ .}
\end{equation}

For DSA the injection  index is given by
\begin{equation}
s_{\rm{inj,dsa}} =  2 \frac{\mathcal{M}^2 +1} {\mathcal{M}^2 -1} \mbox{ .}
\label{eq:inj-mach}
\end{equation}

The distribution after re-acceleration can now be described as follows, if $s_{\rm{fossil}} <  s_{\rm{inj,dsa}}$ then $ s_{\rm{inj,re}} = s_{\rm{fossil}} $. Thus for weak shocks, or a flat distribution of fossil plasma, the shape of the radio spectrum will be preserved under re-acceleration. If $s_{\rm{fossil}} >  s_{\rm{inj,dsa}} $ we have $ s_{\rm{inj,re}} = s_{\rm{inj,dsa}}$, so spectral shape is what we would normally expect from DSA. Note that the radio spectral index is related to electron momentum distribution (with index $s$) as $\alpha = -(s-1)/2$. In summary, for re-acceleration the index of the momentum distribution is given by
\begin{equation}
  s_{\rm{inj,re}}=%
  \begin{cases}
   s_{\rm{fossil}} & \text{for }  \,  s_{\rm{fossil}} <   2 \frac{\mathcal{M}^2 +1} {\mathcal{M}^2 -1}
    \\
 s_{\rm{inj,dsa}}  & \text{for } \, s_{\rm{fossil}} >   2 \frac{\mathcal{M}^2 +1} {\mathcal{M}^2 -1}    \mbox{ .}
  \end{cases}
  \label{eq:proffit}
\end{equation}

At the location where the NAT source in MACS~J0717.5+3745 fades into the relic, the spectral index is  steep with $\alpha \lesssim -2$ ($s\gtrsim5$) and that would suggest that we are in the regime  $s_{\rm{fossil}} >  s_{\rm{inj,dsa}} $ and the spectral index of the relic follows what would be expected in the case of  DSA.  If the spectral index is set by the Mach number, we would need at least a shock with $\mathcal{M} = 2.7$ ($\alpha_{\rm{inj}} = -0.8$). The (current) X-ray observations do not allow us to measure the Mach number, but given the very high gas temperatures, the presence of a shock with $\mathcal{M} \gtrsim 2.7$ cannot be excluded. On the other hand, a $\mathcal{M} = 2.7$ shock would correspond to a factor $\sim 8$ increase in the surface brightness, which should be detectable (unless the shock surface has a very complex shape). Alternatively, we are not in the ``DSA regime'' and the shock has a lower Mach number and is therefore more difficult to detect.

\subsubsection{Shape of the fossil electron distribution before re-acceleration}

\begin{figure}
  \includegraphics[angle=180,width=\columnwidth]{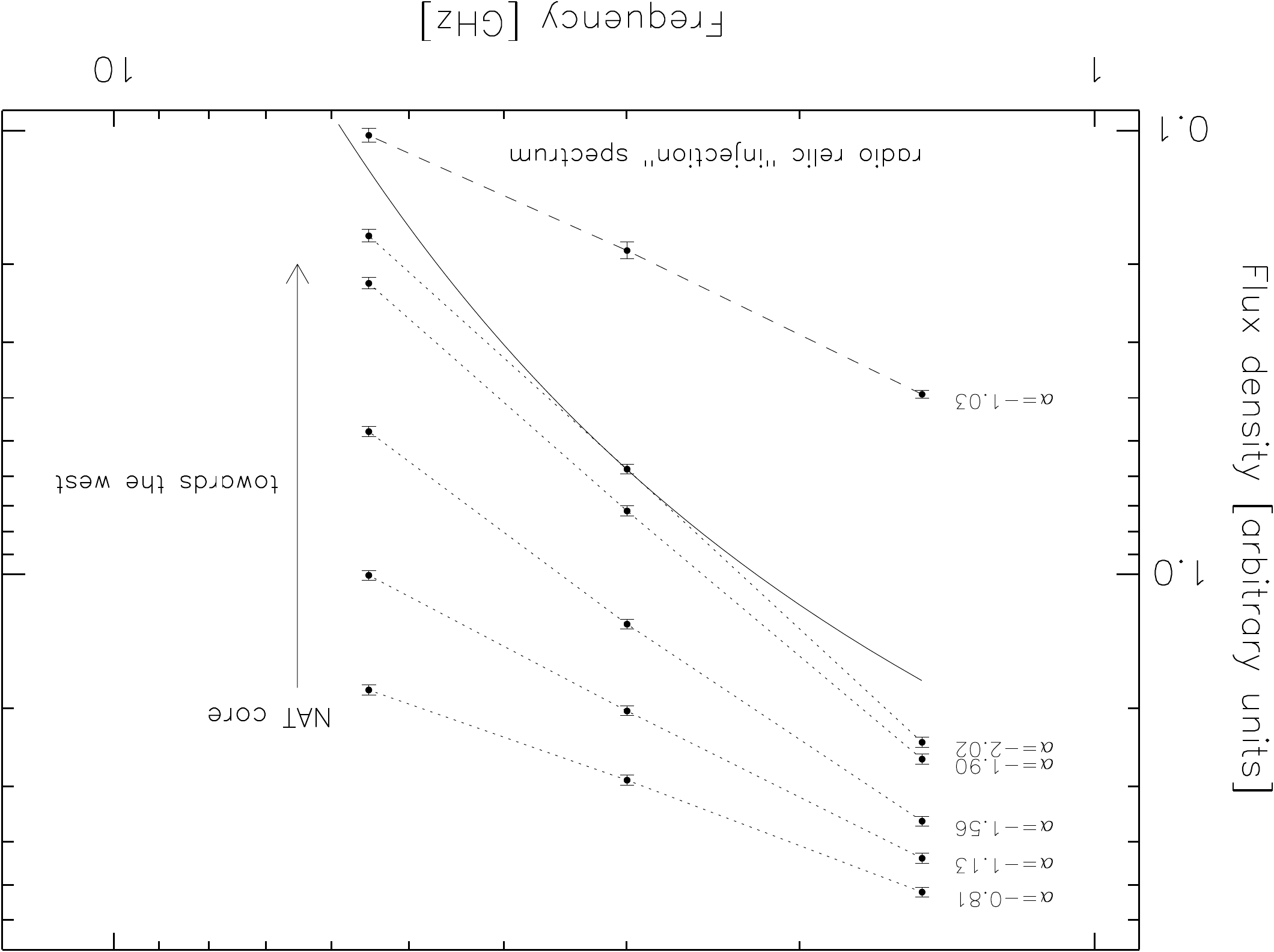}
  \caption{Radio spectra across the NAT source at the center of the relic. The normalizations for all spectra are arbitrary.
  The  upper five spectra are extracted in the magenta regions indicated in Figure~\ref{fig:spixregions}, starting at the easternmost region. The dotted lines are to guide the eye and connect the flux density measurements at 1.5, 3.0 and 5.5~GHz. The solid black line corresponds to a power-law model with an exponential cutoff with $\nu_{\rm{break}} = 2$~GHz. The model spectrum is normalized at our 3.0~GHz flux density measurement. The bottom spectrum, represented by the dashed line, was computed by combining the fluxes from the easternmost relic regions with black and red colors (see Figure~\ref{fig:spixregions}). These parts of the relic contain the flattest spectral indices and thus likely come closest to resembling the radio relic ``injection spectrum''. Spectral index values, between 1,5 and 5.5~GHz, are shown to the left of the spectra.}
\label{fig:NATspectrum}
\end{figure}

Above, we assumed that the fossil electron distribution is that of a power-law.  A more realistic  fossil electron distribution is that of a spectrum that has undergone synchrotron and IC losses. This would change the resulting distribution after re-acceleration from a simple power-law \citep{2015ApJ...809..186K,2015ApJ...812...49H,2016ApJ...823...13K}. Our radio observations allow us to measure the shape of the electron fossil distribution, see Figure~\ref{fig:NATspectrum}.

According to \citet{2015ApJ...809..186K}, the spectrum can be approximated by a power law distribution with some exponential cutoff at frequency $\nu_{\rm{break}}$. We fix the injection spectral index to $\alpha_{\rm{inj}} = -0.5$. The observed spectra do not show evidence for a strong spectral break. In Figure~\ref{fig:NATspectrum} we plot a model with $\nu_{\rm{break}} = 2$~GHz. This lack of a spectral cutoff indicates that the spectral ageing is more complex (e.g., spatially varying magnetic fields) or that there is mixing of radio emission with different spectra within our measurement regions. This mixing reduces the curvature and moves the spectra closer to power-law shapes \citep[e.g.,][]{2012A&A...546A.124V}. Therefore the curvature of the fossil particle spectrum remains unclear, but we can at least conclude that the spectrum is steep. The shape of the fossil distribution will be important input for future modeling and simulations \citep[e.g.,][]{2015ApJ...809..186K,2015ApJ...812...49H}.

 \subsection{Origin of the radio halo and filamentary structures}
An interesting question concerns the origin and nature of the radio filaments in the general radio halo area. Are these embedded in the radio halo emission, tracing regions with increased turbulence, or are they similar to the large relics that trace (re)-accelerated particles at shocks? In the second scenario an additional question is whether they trace shocks in the denser regions of the ICM, or shocks in the cluster outskirts (and in which case they can be projected onto the cluster center and radio halo region). The filaments could also only be regions of enhanced magnetic fields, i.e. flux tubes or large-scale strands of field.

It seems that a shock-origin is preferred, at least for some of the filaments. There are severe reasons for this (1) the northern filament (above R1), which is located in the cluster outskirts, shows an EW spectral index gradient, and has has a well defined eastern boundary; and (2) the filament above R4 is connected with the main radio relic. So at least two of these filaments are probably not directly associated with the radio halo. In addition, it is possible that the regions indicated with the white dashed-line circles (Figure~\ref{fig:3res}) are additional filaments but projected closer to face-on. However, they could also just be regions of enhanced magnetic fields.

Polarization measurements would provide additional information on the filaments. A high ($\gtrsim 20\%$) polarization fraction would indicate aligned magnetic fields due to shock compression. Measurements of the Faraday dispersion function (and Rotation Measure) would allow us to check if the filaments are located in the cluster outskirts on the nearside. Filaments located deep inside the ICM should show higher Faraday
Rotation, although this would also be the case for filaments located on the far side of the cluster. We defer a polarization analysis for this cluster to future work.

\subsection{Merger scenario}

MACS~J0717.5+3745 consists of at least four merging subclusters, as indicated in Figure~\ref{fig:chandraoptical}. Subcluster B, corresponding to the V-shaped structure in the \chandra\ images, has a large line of sight velocity of about 3,200 km~s$^{-1}$ away from us. We speculate that the V-shape could be related to a bullet-like structure seen under a large projection angle. This would also explain the lack of an offset between the X-ray gas, dark matter, and galaxies, and is consistent with the large radial velocity component and the detected kinetic SZ signal \citep{2012ApJ...761...47M,2013ApJ...778...52S}.
Interestingly, the radio filament above R4 is aligned with the V-shape and is located immediately to the south of it. This could just be a chance alignment. Another possibility is that this filament traces the shock ahead of subcluster~B.

For  subcluster~D, the galaxy and dark matter peaks are located about 0.4\arcmin~NW from the X-ray peak of the subcluster. 
An offset in this direction would be expected due to the effect of ram pressure on the gas \citep[as also suggested by][]{2009ApJ...693L..56M}, if subcluster~D fell in from the large-scale galaxy filament to the SE. No clear offset,  between the X-ray peak and dark matter peak, is seen for subcluster~C. \cite{2016arXiv160607721A} reported the detection of a kinetic SZ signal from subcluster~C, with an opposite line of sight velocity with respect to subcluster~B.

\citet{2009ApJ...693L..56M} suggested that subcluster~A (the fly-through core) fell in from the NW. However, the detection of an X-ray edge to the NNE, likely a merger related cold front,  suggests that the cluster fell in from the SE and the X-ray gas is moving to the N-NW. This direction would be consistent with infall from the large-scale filament to the southeast. Its elongated shape indicates it is currently in the process of being ram pressure stripped, see Section~\ref{sec:fly-throughcore}. 
 The associated BCG is located  slightly (0.2\arcmin--0.3\arcmin) to the SE of the X-ray peak. This is different from the situation in the bullet Cluster \citep{2006ApJ...648L.109C}, where the galaxies lead the bullet. This could imply that the dark matter and galaxies are already past pericenter and in the ``return phase'' of the merger \citep{2015MNRAS.453.1531N}. Interestingly, the dark matter peak is located even further to the east \citep[as reported by e.g.,][]{2014ApJ...797...48J,2016A&A...588A..99L}, although other lensing models show better agreement between the BGC location and mass surface density \citep[e.g.,][]{2015ApJ...799...12I}. From a core size of about 0.4\arcmin~($\approx 150$~kpc) and a temperature of 6.8~keV, we compute a sound crossing time of $1\times 10^{8}$~yr for the core. If the X-ray emission is indeed displaced from the dark matter the X-ray clump will disperse over the next $\sim10^{8}$~yr (assuming there is no dark matter to hold it together).

\section{Conclusions}
\label{sec:conclusions}

We presented deep \jvla\ and \chandra\ observations of the HST Frontier Fields cluster MACS~J0717.5+3745. The radio and X-ray observations show a complex  merger event, involving multiple subclusters. 
Below we summarize our findings:

\begin{itemize}

\item The X-ray temperature map shows that the eastern part of the cluster is significantly hotter than the western part. 
In the central southeastern part of the cluster the temperatures exceed $\sim 20$~keV. The hot eastern part of the cluster coincides with the location of the radio halo and relic.

\item We find no evidence for the ICM temperatures significantly less than 10 keV that were reported by \cite{2009ApJ...693L..56M}.

\item The NW subcluster displays a ram pressure-stripped core, with a surface brightness edge to the NNE. We speculate that this edge is likely a merger related cold front.

\item We find evidence that the radio relic in MACS~J0717.5+3745 is powered by shock re-acceleration of fossil electrons from a nearby NAT source.

\item We find an overall EW spectral index gradient across the radio relic, with the spectral index steepening towards the west.

\item We do not detect density or temperatures jumps associated with the radio relic, which could be the result of the complex merger geometry. Alternatively, for re-acceleration the shock Mach number could be lower than the $\mathcal{M}=2.7$ calculated from the radio spectral index.

\item We find several radio filaments in the cluster with sizes of about 100--300~kpc. At least a few of these are located in the cluster outskirts. That would suggest the filaments are tracing shock waves (and can thus be classified as a small radio relics). Polarization observations should provide more information about the origin and location of these filaments within the ICM.

\end{itemize}

\acknowledgments
{\it Acknowledgments:}
We thank the anonymous referee for useful comments.
The National Radio Astronomy Observatory is a facility of the National Science Foundation operated under cooperative agreement by Associated Universities, Inc. Support for this work was provided by the National Aeronautics and Space Administration through Chandra Award Number GO4-15129X issued by the Chandra X-ray Observatory Center, which is operated by the Smithsonian Astrophysical Observatory for and on behalf of the National Aeronautics Space Administration under contract NAS8-03060.

R.J.W. is supported by a Clay Fellowship awarded by the Harvard-Smithsonian Center for Astrophysics. M.B acknowledge support by the research group FOR 1254 funded by the Deutsche Forschungsgemeinschaft: ``Magnetisation of interstellar and intergalactic media: the prospects of low-frequency radio observations''. W.R.F., C.J., and F.A-S. acknowledge support from the Smithsonian Institution. E.R. acknowledges a Visiting Scientist Fellowship of the Smithsonian Astrophysical Observatory, and the hospitality of the Center for Astrophysics in Cambridge.  G.A.O. acknowledges support by NASA through a Hubble Fellowship grant HST-HF2-51345.001-A awarded by the Space Telescope Science Institute, which is operated by the Association of Universities
for Research in Astronomy, Incorporated, under NASA contract NAS5-26555.  F.A-S. acknowledges support from Chandra grant GO3-14131X. A.Z. is supported by NASA through Hubble Fellowship grant HST-HF2-51334.001-A awarded by STScI. This research was performed while T.M. held a National Research Council Research Associateship Award at the Naval Research Laboratory (NRL). Basic research in radio astronomy at NRL by T.M. and T.E.C. is supported by 6.1 Base funding. M.D. acknowledges the support of STScI grant 12065.007-A. P.E.J.N. was partially supported by NASA contract NAS8-03060. Part of this work performed under the auspices of the U.S. DOE by LLNL under Contract DE-AC52-07NA27344.

Part of the reported results are based on observations made with the NASA/ESA Hubble Space Telescope, obtained from the Data Archive at the Space Telescope Science Institute. STScI is operated by the Association of Universities for Research in Astronomy, Inc. under NASA contract NAS 5-26555. This work utilizes gravitational lensing models produced by PIs Brada{\v c}, Ebeling, Merten \& Zitrin, Sharon, and Williams funded as part of the HST Frontier Fields program conducted by STScI. The lens models were obtained from the Mikulski Archive for Space Telescopes (MAST).



{\it Facilities:} \facility{VLA}, \facility{CXO} 



\appendix

\section{Spectral index uncertainty maps}
\label{sec:spixerror}

\begin{figure*}[h]
\begin{center}
\includegraphics[angle =180, trim =0cm 0cm 0cm 0cm,width=0.49\columnwidth]{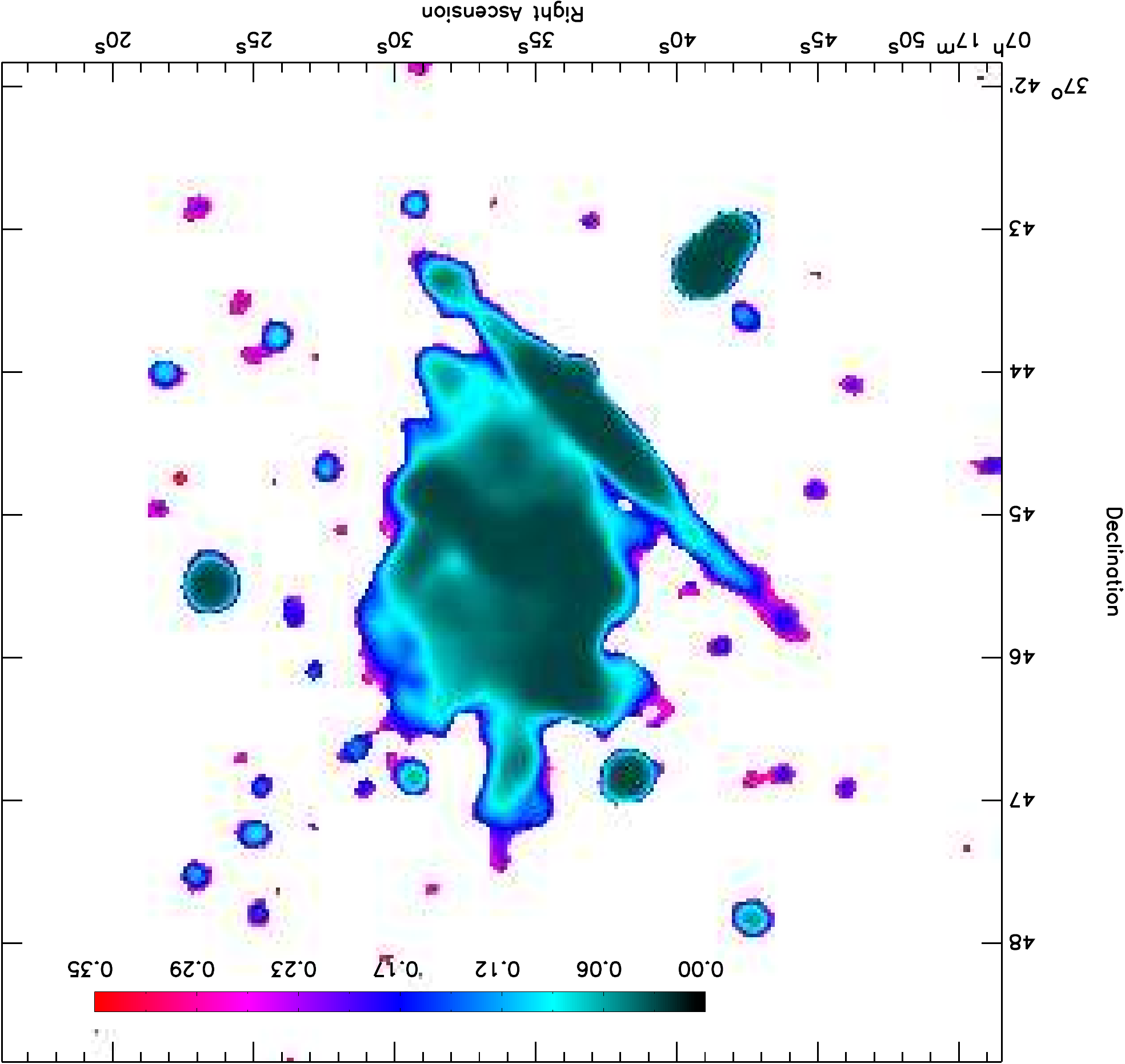}
\includegraphics[angle =180, trim =0cm 0cm 0cm 0cm,width=0.49\columnwidth]{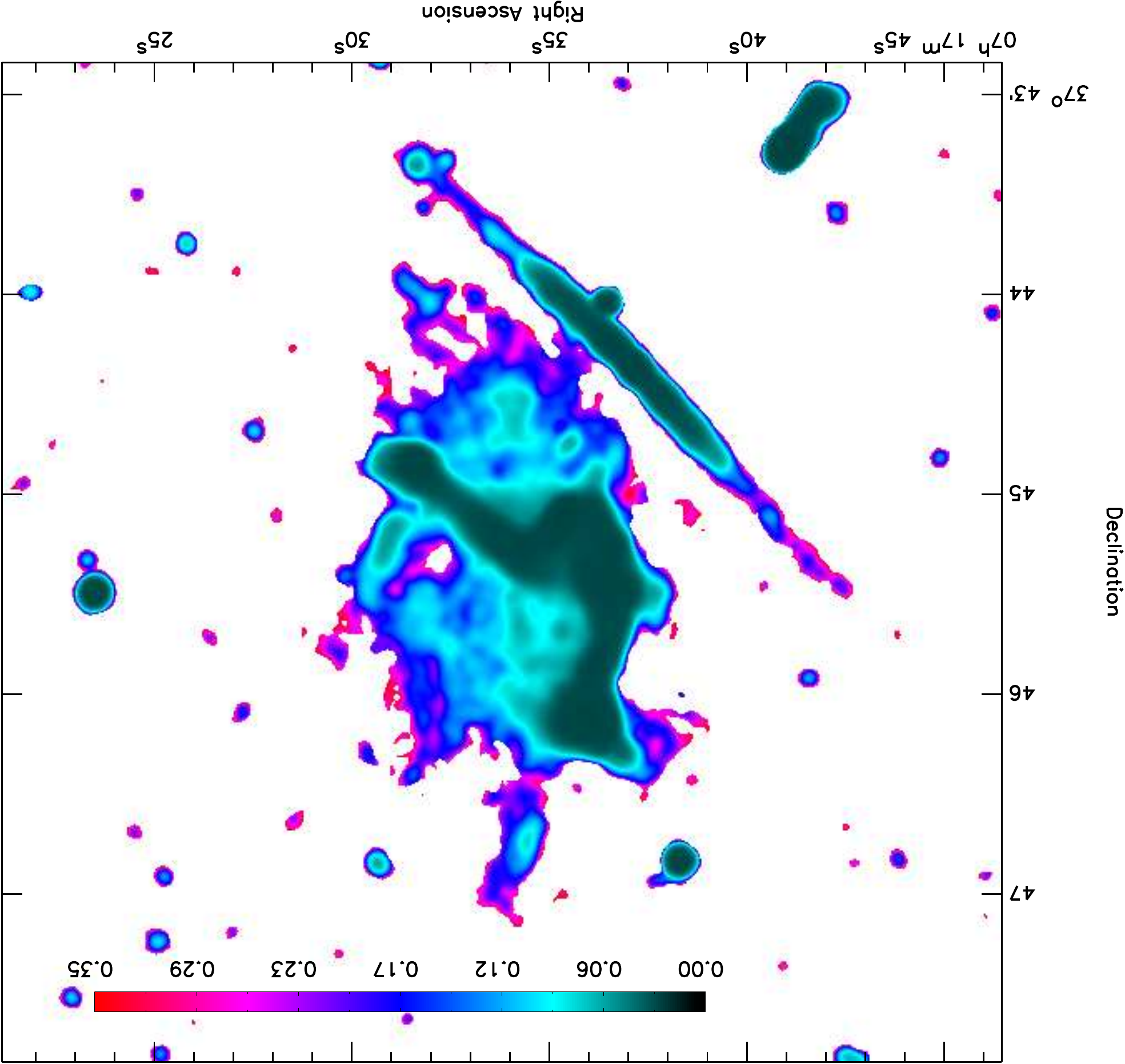}
\includegraphics[angle =180, trim =0cm 0cm 0cm 0cm,width=0.49\columnwidth]{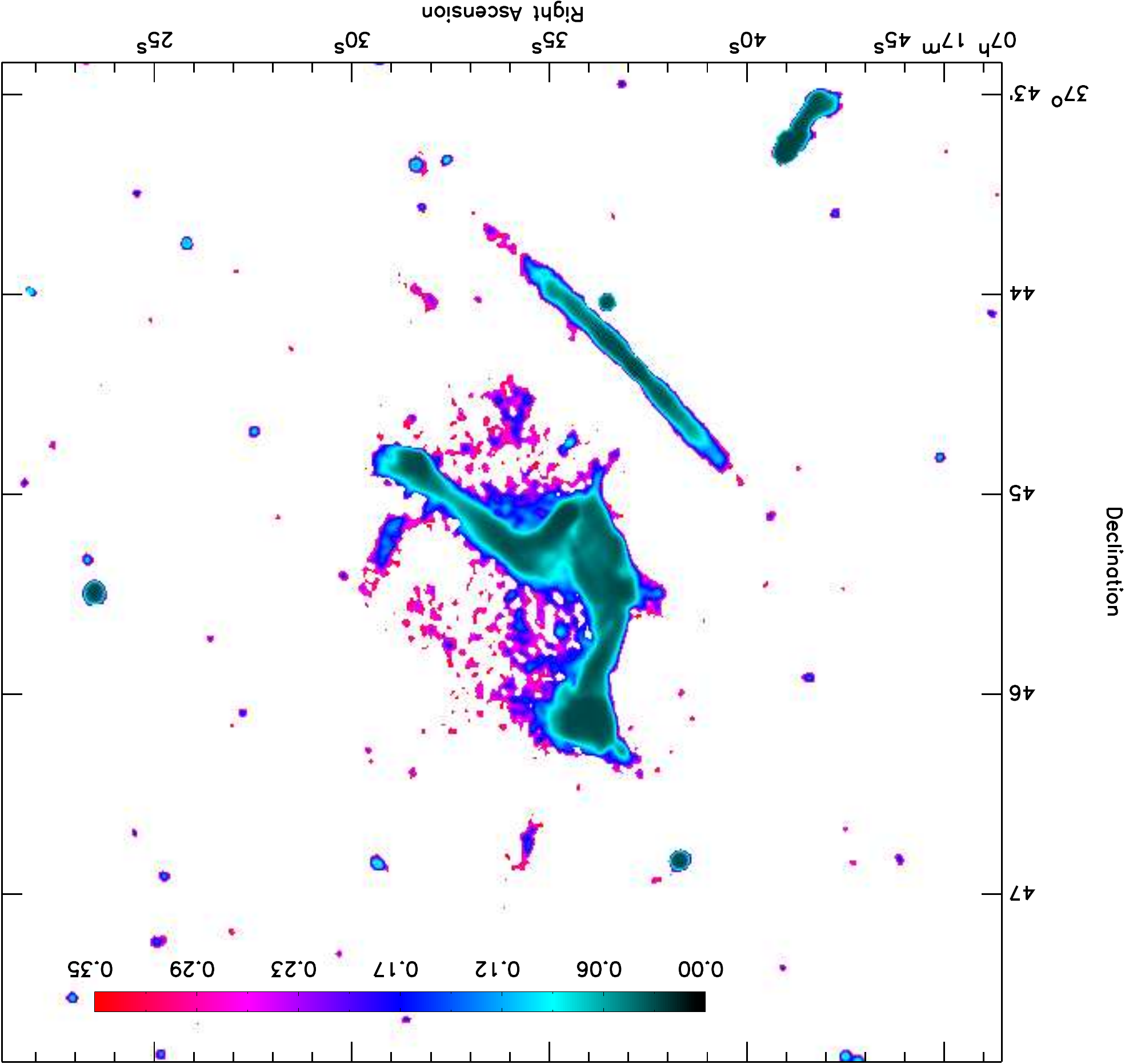}
\includegraphics[angle =180, trim =0cm 0cm 0cm 0cm,width=0.49\columnwidth]{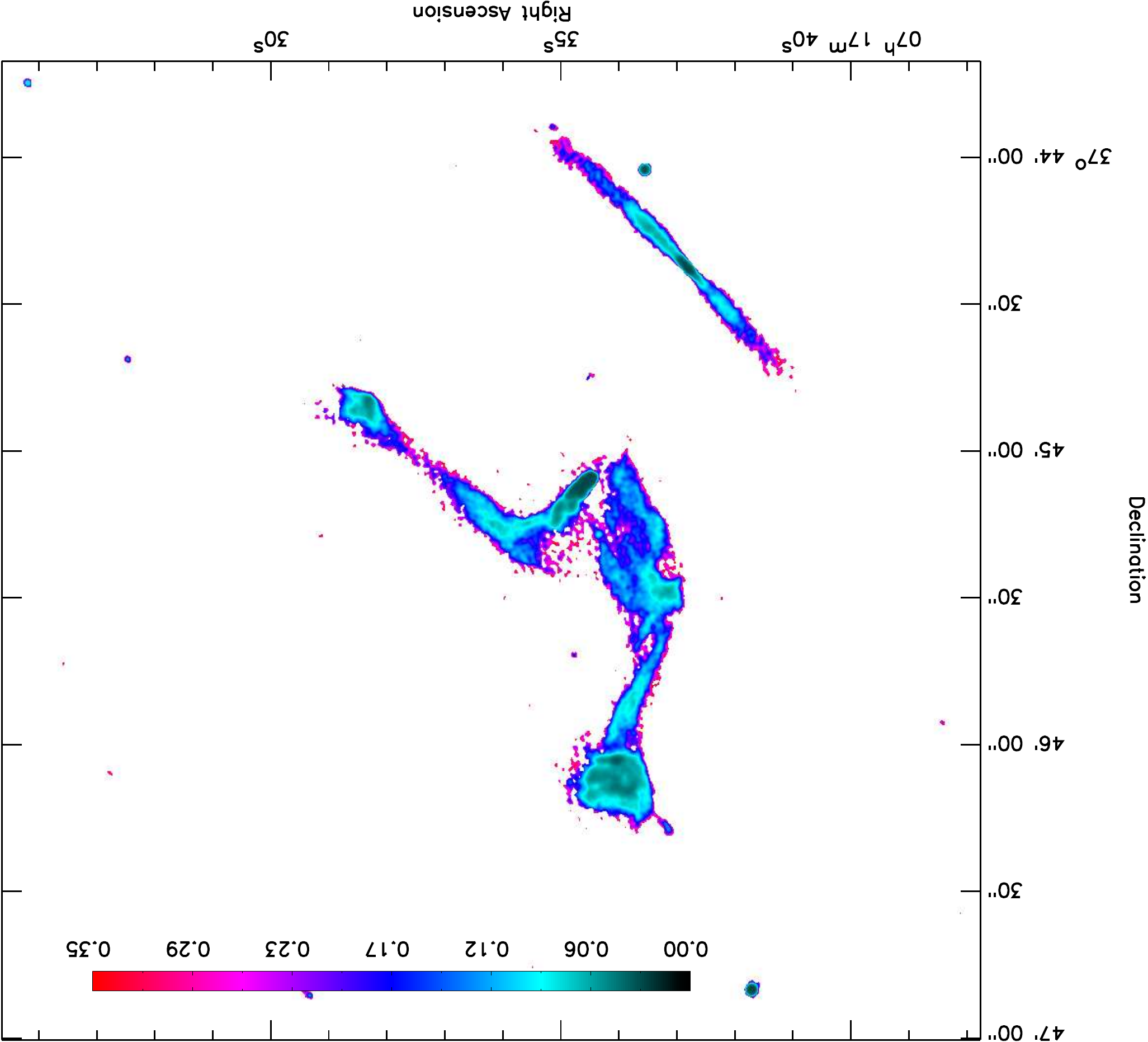}
\end{center}
\caption{Spectral index uncertainty  maps, corresponding to Figure~\ref{fig:spix}, at resolutions of 10\arcsec, 5\arcsec, 2.5\arcsec, and 1.2\arcsec. }
\label{fig:spixu}
\end{figure*}

Spectral index uncertainty maps are shown in Figure~\ref{fig:spixu} corresponding to a power-law fits through flux measurements at 1.5, 3.0 and 5.5~GHz. The errors are based on the individual rms noise values in the maps and an absolute flux calibration uncertainty of 2.5\% at each of the three frequencies. 

\bibliography{ref_filaments}

\end{document}